\documentclass[12pt]{article}%
\usepackage{cite}
\usepackage{fix-cm}
\usepackage{heck}
\usepackage{graphicx}
\usepackage{multicol}
\usepackage{amsmath}
\usepackage{amsfonts}
\usepackage{amssymb}
\usepackage{hyperref}
\usepackage{setspace}
\usepackage[all]{xy}
\usepackage{verbatim}
\usepackage{color}
\usepackage{epsfig}
\usepackage{mathtools}
\providecommand{\U}[1]{\protect\rule{.1in}{.1in}}
\numberwithin{equation}{section}

\def\be{\bea}
\def\ee{\eea}
\def\bea{\begin{eqnarray}}
\def\eea{\end{eqnarray}}
\def\spc{\hspace{1pt}}

\newcommand{\smpc}{\hspace{.5pt}}

\def\is{\! & \! = \! & \! }
\def\ra{\rangle}
\def\la{\langle}
\def\ellcc{\ell}
\def\Dbar{\overline{D}}

\begin{document}


\date{December 2011}

\title{Instantons, Twistors, and Emergent Gravity}

\institution{IAS}{\centerline{${}^{1}$School of Natural Sciences, Institute for Advanced Study, Princeton, NJ 08540, USA}}

\institution{PU}{\centerline{${}^{2}$Department of Physics, Princeton University, Princeton, NJ 08544, USA}}

\authors{Jonathan J. Heckman\worksat{\IAS}\footnote{e-mail: {\tt jheckman@ias.edu}} and Herman Verlinde \worksat{\PU}\footnote{e-mail: {\tt verlinde@princeton.edu}}}

\abstract{Motivated by potential applications to holography on space-times of positive curvature, and by the successful
twistor description of scattering amplitudes, we propose a new dual matrix formulation of  $\mathcal{N} = 4$ gauge theory on $S^4$.
The matrix model is defined by taking the low energy limit of a holomorphic Chern-Simons theory on $\mathbb{CP}^{3|4}$, in the presence of a large instanton flux.
The theory comes with a choice of $S^4$ radius $\ellcc$ and a parameter $N$ controlling the overall size of the matrices.
The flat space variant of the 4D effective theory arises by taking the large $N$ scaling limit of the matrix model, with $\ell_{pl}^{2} \sim \ellcc^{2} / N$ held fixed. Its massless spectrum contains both spin one and spin two excitations, which we identify with gluons and gravitons.
As shown in the companion paper \cite{GMM}, the matrix model correlation functions of both these excitations correctly reproduce the corresponding MHV scattering amplitudes. We present evidence that the scaling limit defines a gravitational theory with a finite Planck length. In particular we find that in the $\ell_{pl} \to 0$ limit, the matrix model makes contact with the CSW rules for amplitudes of pure gauge theory, which are uncontaminated by conformal supergravity.
We also propose a UV completion for the system by embedding the matrix model
in the physical superstring.}

\maketitle

\enlargethispage{\baselineskip}

\setcounter{tocdepth}{2}
\tableofcontents

\newpage

\renewcommand\Large{\fontsize{16.5}{18}\selectfont}
\section{Introduction}

While there is little doubt that string theory provides a consistent theory of
quantum gravity, it has proven rather difficult to specify the physical foundations
of the theory. Part of the issue is that in situations where maximal theoretical control
is available, space-time is treated as a classical background, rather than as an emergent
concept. Related to this, the understanding of holography on space-times of positive curvature
remains elusive.

In this paper we propose and develop a new dual matrix formulation of 4D field
theory, in which the space-time and field theory degrees of freedom simultaneously emerge
from a large $N$ double scaling limit. The basic idea is to view ${\cal N}\! =\! 4$  gauge theory
on $S^{4}$ as an effective low energy description of an underlying topological large  $N$
gauge theory, without  any local propagating degrees of freedom.
This topological theory takes the form of  holomorphic Chern-Simons (hCS) theory defined on
projective supertwistor space $\mathbb{CP}^{3|4}$. Motivated by the (partial)
successes of earlier matrix reformulations of string theory \cite{Banks:1996vh, Ishibashi:1996xs, juanAdS, gkPol, witHolOne},
and of the twistor string reformulation of gauge theory \cite{Witten:2003nn}, we consider the
hCS theory in the presence of a large background flux \cite{TMM}. We
choose this flux to be a homogeneous $U(N)$ instanton on $S^4$ with the maximal
possible instanton number $k_N$.   This special flux configuration, known as the $U(N)$ Yang monopole,\footnote{The
Yang monopole was actively studied in the context of a proposed 4D realization of the Quantum
Hall effect in \cite{Zhang:2001xs}. These authors also found emergent spin 2 modes among the edge
excitations of the QH droplets, though there are some issues with interpreting these modes
as connected with an emergent gravitational sector (see e.g. \cite{Elvang:2002jh}). Our setup has some similarities with the system of \cite{Zhang:2001xs}, but differs in several essential aspects.} breaks the $U(NN_c)$ gauge symmetry of the UV theory
down to $U(N_c)$, the gauge group of the IR effective gauge theory.

As we will show, the low energy physics in the presence of the flux is governed by a large $N$ matrix model,
with matrices of size $k_N \times k_N$. The guiding principle that determines the form and
content of the matrix model action
is the requirement that the matrix equations of motion encapsulate the space of deformations of
the Yang monopole, and reduce to the ADHM equations in the appropriate flat space limit.
The finite matrices arise because, in the presence of the instanton flux,  all charged
excitations are forced to move in Landau levels. The lowest level is built up from $k_N$ Planck
cells -- or fuzzy points -- of a non-commutative $\mathbb{CP}^{3|4}$, where each fuzzy
point represents one elemental instanton. The coordinates of the supermanifold become operators, that satisfy a commutator algebra
\bea
\label{een}
\lbrack Z^\alpha ,Z_\beta^{\dag}]=\delta^{\alpha}_{\, \beta} . 
\eea
The oscillator number is bounded to be less than some integer  $N$, which translates
into an upper limit on the allowed angular momentum on the $S^{4}$.

A striking aspect of this type of non-commutative deformation is that, unlike other more standard versions
of space-time non-commutativity, it preserves all space-time isometries. Moreover, because the
non-commutative deformation does not directly affect the holomorphic coordinates,
 the basic geometric link between the twistor variables and 4D physics is kept intact.
Nevertheless, the instanton density of the background flux introduces a UV length scale on the $S^{4}$,
which in units of the $S^4$ radius $\ellcc$ scales as
\bea
\label{elpee}
\ell_{pl}^{2}\simeq \frac{\ell^{2}}{N}\, .
\eea
Below this length scale, the physics is described by a holomorphic Chern-Simons theory,
expanded around a trivial background and without local excitations. At much larger scales, however, the
collective motion of the instanton background produces rich dynamics that, as we will show,
takes the form of a quantum field theory with a UV cutoff  given by (\ref{elpee}).
The continuum QFT arises by taking the large $N$ limit while keeping the $S^4$ radius $\ellcc$ fixed.
 One can instead take a combined large $N$ and flat space limit, by simultaneously sending $N$
 and $\ell$ to infinity while keeping $\ell_{pl}$
fixed. As we will argue, this results in a 4D gravitational theory with a finite Planck length.

The emergence of gravity from an a priori non-gravitational dual system is
a remarkable phenomenon that deserves further exploration. At a superficial level,
our matrix model looks like a regulated version of the twistor string theory introduced by Witten
as a dual description of ${\cal N}\! = \! 4$ gauge theory.
Twistor string theory also gives rise to gravitational degrees of freedom in the form of
conformal supergravity. However, our set-up differs in several essential aspects from the
twistor string. Most importantly, our theory comes with a natural length scale that breaks conformal
invariance. Moreover, the emergence of gravity has a clear geometric origin via
the non-commutativity of twistor space.

The non-commutativity (\ref{een}) of the twistor coordinates causes gauge transformations
to act like diffeomorphisms. Correspondingly, a non-zero gauge field background on
non-commutative twistor space translates into a deformation of the emergent 4D space-time
geometry. This relationship can be made precise, via a direct variant of Penrose's
non-linear graviton construction of self-dual solutions to the 4D Einstein equations.
In a companion paper \cite{GMM} we use this insight to evaluate correlation
functions of bilinear matrix observables,
and show that quite remarkably, the end result reproduces MHV graviton scattering amplitudes
of ordinary 4D  Einstein (super)gravity.

\begin{figure}[ptb]
\begin{center}
\includegraphics[
height=2.0349in,
width=2.0928in
]{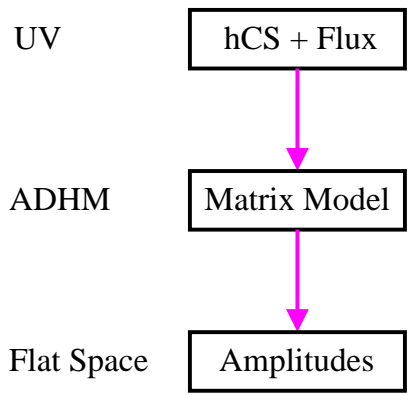}
\end{center}
\caption{The twistor matrix model captures the low energy physics of
holomorphic Chern-Simons theory on $\mathbb{CP}^{3|4}$ in the presence of an instanton background.
The ingredients of the matrix model have a natural connection with the ADHM\ construction of instantons. Correlation functions of the
matrix model compute amplitudes of a 4D space-time theory, which in the flat space limit  coincide with scattering amplitudes
of gluons and gravitons.}%
\label{flowchart}%
\end{figure}

The matrix model is the low energy approximation to a more complete UV theory.
In the first part of the paper, the UV theory is taken to be the hCS theory, or topological B-model open string theory,
defined on super twistor space. In the last section, we propose an alternative UV completion  in terms of
type IIA superstring theory. The brane configuration that we consider is given by a D0-brane bound
to a stack of D8-branes and an O8-plane. The background geometry is taken to be 10 non-compact space-time
dimensions. In the presence of a background magnetic flux, the low temperature limit of a gas of fermionic 0-8
strings reproduces the degrees of freedom and action of the twistor matrix model. An intriguing aspect of this
construction is that it (potentially) gives rise to a 4D theory coupled to gravity, without first having to choose
some Kaluza-Klein compactification manifold.

The rest of this paper is organized as follows. In section \ref{sec:PRELIM} we
review some basics of twistor geometry. Next, in
section \ref{sec:FUZZ} we introduce holomorphic Chern-Simons theory
with the Yang monopole flux and explain how it naturally gives rise to a non-commutative
deformation of twistor space. This is followed in section
\ref{sec:TMM} by the presentation of the twistor matrix model, and the gauge
currents of the defect system. Section \ref{sec:ADHM} is devoted to matching
a particular limiting case of the matrix model to the ADHM construction of instantons.
We study the continuum limit of the matrix model
in section \ref{sec:LRAD}. Section \ref{sec:GRAVFUZZ} discusses the sense in
which the theory describes a quantum theory of gravity, and in section \ref{Embedder}
we provide a UV completion of the matrix model based on thermodynamics of a
stringy brane construction. We present our conclusions and avenues of future
investigation in section \ref{sec:CONC}.

\section{Twistor Preliminaries \label{sec:PRELIM}}

In this section we review some preliminary aspects of twistor
geometry, as well as the connection between hCS and self-dual Yang Mills. We
also discuss quantization of twistor space.

\subsection{Twistor Correspondence}

Twistors were introduced by Penrose as a natural geometric description of
massless field theories \cite{Penrose:1967wn, Penrose:1968me} (see
e.g. \cite{Penrose:1986ca, WardWells} for reviews). The construction starts with the introduction of spinor
helicity variables. In complexified Minkowski space, light-like momenta
$p_{a\dot{a}}$ can be represented as a product $p_{a\dot{a}}=\widetilde{\pi
}_{\dot{a}}\pi_{a}$ of a chiral and an anti-chiral spinor. Next, one notices
that given the spinor $\pi_{a}$ and a space-time point $x^{a\dot{a}}$ on
complexified Minkowski space, one can define a corresponding two component
complex spinor $\omega^{\dot{a}}$ via
\bea
\omega^{\dot{a}}\is ix^{\dot{a}a}\pi_{a}. \label{twistorequation}%
\eea
This relation is invariant under simultaneous complex rescaling of the two
spinors $\pi_{a}$ and $\omega^{\dot{a}}$. One thus obtains a set of
homogeneous coordinates $Z^{\alpha}=(\omega^{\dot{a}},\pi_{a})$ on projective
twistor space $\mathbb{PT}^{\bullet}$. Similarly, one can introduce a dual
twistor space, with coordinates $\widetilde{Z}_{\beta}=(\widetilde{\pi}%
_{\dot{a}},\widetilde{\omega}^{a})$, denoted by $\mathbb{PT}_{\bullet}$.
Geometrically, $\mathbb{PT}_{\bullet}$ and its dual both define a
three-dimensional complex projective space $\mathbb{CP}^{3}$. We will raise
and lower two component spinor indices with the help of the corresponding two
component epsilon symbols.\footnote{Our conventions are as follows:
$\omega_{\dot{b}}=\omega^{\dot{a}}\varepsilon_{\dot{a}\dot{b}}$, $\pi^{b}%
=\pi_{a}\varepsilon^{ab}$.} The associated antisymmetric spinor inner products
are denoted via
\bea
\langle\pi_{1}\pi_{2}\rangle=\varepsilon^{ab}\pi_{1a}\pi_{2b},\quad &  &
\quad\lbrack\omega_{1}\omega_{2}]=\varepsilon_{\dot{a}\dot{b}}\omega_{1}%
^{\dot{a}}\omega_{2}^{\dot{b}}\,. \label{spinpair}%
\eea
Equation (\ref{twistorequation}) combines two complex linear relations in
$\mathbb{CP}^{3}$. For a given point $x$ in complexified space-time, it
defines a $\mathbb{CP}^{1}$ , which we will call the \emph{twistor line}
associated with $x$. The space $\mathbb{C}^{4}\times\mathbb{CP}^{1}$ with
coordinates $(x,\pi)$ is known as the \textquotedblleft correspondence
space\textquotedblright. This admits a map to $\mathbb{CP}^{3}$ via
$\omega^{\dot{a}}=ix^{\dot{a}a}\pi_{a}$.

A convenient presentation of complexified, conformally compactified Minkowski
space is given as the zero locus of the Klein quadric in $\mathbb{CP}^{5}$:
\bea
\varepsilon_{\alpha\beta\gamma\delta}X^{\alpha\beta}X^{\gamma\delta}\is 0\ .
\label{KleinQ}%
\eea
Here the $X^{\alpha\beta}=-X^{\beta\alpha}$ are the six homogeneous
coordinates of $\mathbb{CP}^{5}$.
The constraint (\ref{KleinQ}) is automatically solved by introducing a pair of
points in twistor space, with homogeneous coordinates $Z^{\alpha}$ and
$W^{\beta}$, via%
\bea
X^{\alpha\beta}\is Z^{[\alpha}W^{\beta]}\,. \label{twisquadX}%
\eea
These homogeneous coordinates satisfy: $X^{[\alpha\beta}Z^{\gamma]}=0$.
For a given $X^{\alpha\beta}$ these equations are solved by all points on the line
 through $Z$ and $W$.

\def\xX{{\mbox{\fontsize{13.5}{13}\selectfont $x$}}}

Since the $X^{\alpha\beta}$ are homogeneous coordinates, they are only
sensitive to the conformal structure of space-time. Conformal symmetry is
broken by designating a choice of two index anti-symmetric bi-twistor, called
the \emph{infinity twistor}, which we denote by $I_{\alpha\beta}%
=-I_{\beta\alpha}$. The `inverse' bi-twistor is denoted by $I^{\alpha\beta
}=\frac{1}{2}\varepsilon^{\alpha\beta\gamma\delta}I_{\gamma\delta}$. The
infinity twistor allows us to raise and lower the index of the twistor
coordinates $Z^{\alpha}$ via $Z_{\alpha}\equiv I_{\alpha\beta}Z^{\beta}$.
Furthermore, it defines an anti-symmetric pairing $\langle\hspace
{1pt}ZW\rangle$ between two twistors $Z$ and $W$ via
\bea
\langle\spc ZW\rangle\is I_{\alpha\beta}Z^{\alpha}W^{\beta}%
\eea
As the name suggests, the infinity twistor selects the twistor lines that map
to space-time points at infinity: a pair $ZW$ describes an infinite point if
$\langle\hspace{1pt}ZW\rangle=0$. Deleting the locus $\langle\hspace
{1pt}ZW\rangle=0$, we can introduce affine coordinates
$\xX^{\alpha\beta}$, and their duals
$\tilde\xX_{\alpha\beta}$, via\footnote{Note that pairs $ZW$
with $\langle ZW\rangle=0$ indeed map to points in complexified space-time
with $\xX^{\alpha\beta}=\infty$. For $S^{4}$, this
infinity locus does not intersect with the real four sphere. In the flat space
limit, $\ell\rightarrow\infty$, the infinity twistor degenerates, and selects
$\pi_{a}=0$ as the twistor line associated with space-like infinity.}
\bea
\xX^{\alpha\beta}=\frac{Z^{[\alpha}W^{\beta]}}%
{\langle\hspace{1pt}ZW\rangle},\quad & & \quad\tilde\xX_{\alpha\beta}=\frac{1}{2}\epsilon_{\alpha\beta\gamma\delta}%
\xX^{\gamma\delta} \label{Xhatted}%
\eea
Raising and lowering of a single index is accomplished via the infinity
bitwistor. These coordinates satisfy the relations $\xX^{\alpha\gamma}\xX_{\gamma}{}^{\beta}=-\xX^{\alpha\beta},$ and $\tilde\xX_{\alpha\gamma}\spc
\tilde\xX^{\gamma}{}_{\beta}=-\tilde{\xX}_{\alpha\beta}$.
The dual coordinates $\tilde\xX_{\alpha\beta}$ act as
projection matrices onto the twistor line associated to the space-time point
${\mbox{\small $\hat{X}$}}$:%
\bea
\tilde{\xX}_{\alpha\beta}{U}^{\beta}\is 0
\label{projequation}%
\eea
which is solved by all points $U=a\hspace{1pt}Z+b\hspace{1pt}W$ on the twistor
line through $Z$ and $W$.

The  form of the infinity twistor depends on the choice of space-time signature and metric.
In most of this paper, we will work in Euclidean signature.
Euclidean twistor space can be obtained from
$\mathbb{CP}^{3}$ by introducing a map $\sigma$ which acts on the coordinates
$Z^{\alpha}$ as:%
\bea
\left(  Z^{1},Z^{2},Z^{3},Z^{4}\right) &  \overset{\sigma}{\rightarrow} & \left(
\overline{Z}^{2},-\overline{Z}^{1},\overline{Z}^{4},-\overline{Z}^{3}\right)
\eea
where $\overline{Z}^{\alpha}$ denotes the complex conjugate of
$Z^{\alpha}$. The point $Z^{\alpha}$ and its image $W^\alpha = \sigma\left(  Z^{\alpha
}\right)  $ define a corresponding
space-time point via equation (\ref{twisquadX}). None of the resulting twistor
lines intersect. This is why $\mathbb{CP}^{3}$ can be described as an $S^{2}$
fibered over $S^{4}$.

The infinity twistor of the euclidean four sphere of radius $\ell$ can be chosen of the form
\bea
I_{\alpha\beta}\is \frac{1}{\ellcc}%
\begin{pmatrix}
\hspace{1pt}\gamma\varepsilon_{\dot{a}\dot{b}} & 0\\[1mm]%
0 & \varepsilon^{ab}%
\end{pmatrix}
\ . \label{infinitytwistor}%
\eea
When $\gamma=1$, we have a round $S^{4}$ with isometry group $SO(5)$. When
$\gamma=0$, we have the infinity twistor for Minkowski space. Decomposing the
twistor into two component spinors
\bea
Z^{\alpha}=  (\hspace{1pt}\omega^{\dot{a}},\pi_{a}), & & {Z}_{\alpha}=\frac
{1}{\ellcc}\left(  \!\!%
\begin{array}
[c]{c}%
\hspace{1pt}\gamma\omega_{\dot{a}}\\
\pi^{a}%
\end{array}
\!\!\!\right)  \,,
\eea
the twistor inner product takes the form
\bea
\langle Z_{1}Z_{2}\rangle\is \frac{1}{\ellcc}\langle\pi_{1}\pi_{2}%
\rangle+\hspace{1pt}\frac{\gamma}{\ellcc}[\omega_{1}\omega_{2}]\, ,
\eea
with $\langle\,..\,\rangle$ and $[\,..\,]$ the spinor pairings (\ref{spinpair}%
). Note that, unlike for Minkowski space, the infinity twistor
(\ref{infinitytwistor}) for non-zero $\gamma$ is invertible.

The correspondence between the two twistor line equations
(\ref{twistorequation}) and (\ref{projequation}) is made as follows. The dual
space-time coordinates $\tilde\xX_{\;\beta
}^{\hspace{1pt}\alpha}$ can be parameterized with the help of five coordinates
$y^{A}$ constrained to live on a round $S^{4}$ of
unit radius, via%
\bea
\qquad \xX_{\alpha\beta}\is %
\begin{pmatrix}
\hspace{1pt}\frac{1}{2}(1+ y_{5})\epsilon^{ab}\, &
-iy^{a}{}_{\dot{b}}\,\\[2mm]%
\,{i}y_{\dot{a}}{}^{b}\, & \frac{1}{2}%
(1-y_{5})\epsilon_{\dot{a}\dot{b}}%
\end{pmatrix}
,\qquad\quad y^{A}y_{A}=1\,.
\label{xparam}%
\eea
Here $y_{b}^{\;\dot{a}}=\frac{1}{2}%
y_{\mu}(\sigma^{\mu}){}_{b}^{\;\dot{a}}$ with
$\sigma^{\mu}=(\sigma^{i},-i\mathbf{1})$ the usual Pauli matrices. With this
parametrization, the four component twistor line equation (\ref{projequation})
reduces to the standard two-component twistor line equation
(\ref{twistorequation}) with
\bea
{x_{b}^{\;\dot{a}}}\is \,\frac{2 y_{b}%
^{\;\dot{a}}}{1-y^{5}}\,.
\eea
The flat space limit amounts to zooming in on the south pole region of the
$S^{4}$ near $y_5 \simeq-1$. In this
limit, the remaining four $S^{4}$ coordinates $y^{\mu
}$ become identified with the flat space coordinates $x^{\mu}$.

The geometry of twistor space also extends to theories with supersymmetry.
Here we focus on theories with $\mathcal{N}=4$ supersymmetry. This is achieved
by supplementing the four bosonic coordinates by four fermionic coordinates
$\psi^{i}$ $i=1,...,4$. The resulting supertwistor space $\mathbb{CP}^{3|4}$
is a Calabi-Yau supermanifold. The homogeneous coordinates of $\mathbb{CP}^{3|4}$
will be denoted as $\mathcal{Z}^{I}=(Z^{\alpha}|\psi^{i})$. Chiral Minkowski
superspace is then described by coordinates $(x_{-}^{\dot{a}a},\theta_{-}%
^{ia}).$ One can also introduce coordinates on Minkowski superspace given as
$(x_{+}^{\dot{a}a},x_{-}^{\dot{a}a},\theta_{+i}^{\dot{a}},\theta_{-}^{ia})$
subject to the constraint equations:%
\bea
x_{\pm}^{\dot{a}a}\is x^{\dot{a}a}\pm\theta_{+i}^{\dot{a}}\theta_{-}^{ia}%
\eea
so that the independent coordinates on complexified superspacetime
$\mathbb{M}^{4|16}$ are $(x,\theta_{+i}^{\dot{a}},\theta_{-}^{ia})$. Chiral
and anti-chiral superspace are then parameterized by $(x_{-}^{\dot{a}a},\theta_{-}^{ia})$ and
$(x_{+}^{\dot{a}a},\theta_{+i}^{\dot{a}})$, respectively. Let us note
that in the complexified setting, $\theta_{+}$ and $\theta_{-}$ constitute
independent coordinates. A bosonic line in
supertwistor space is specified by:%
\bea
\omega^{\dot{a}}=ix_{-}^{\dot{a}a}\pi_{a} & ; & \psi^{i}=\theta_{-}%
^{ia}\pi_{a} \label{bosonicline}%
\eea
which are invariant under re-scalings of the superspace coordinate
$\mathcal{Z}^{I}$, and so define homogeneous coordinates on projective
supertwistor space $\mathbb{CP}^{3|4}$. Equations (\ref{bosonicline}) define a
bosonic $\mathbb{CP}^{1|0}$ in $\mathbb{CP}^{3|4}$. The correspondence space
is now parameterized by coordinates $(x,\theta_{-},\pi)$ for $\mathbb{C}%
^{4|8}\times\mathbb{CP}^{1|0}$, with the map to $\mathbb{CP}^{3|4}$ specified
by equations (\ref{bosonicline}).

There is a natural extension of the infinity twistor to a super bi-twistor,
$\mathcal{I}_{IJ}$. In the case of $S^{4|8}$, the super bi-twistor is:%
\bea
\label{supinftwist}
\mathcal{I}_{IJ}\is \frac{1}{\ellcc}\left(  \!%
\begin{array}
[c]{ccc}%
\gamma\varepsilon_{\dot{a}\dot{b}} & 0 & 0\\
0 & \varepsilon^{ab} & 0\\
0 & 0 & \eta_{ij}\!
\end{array}
\right)
\eea
where $\eta_{ij}$ is a symmetric tensor.

\subsection{Holomorphic Chern-Simons}

Two of the most notable successes of twistor theory are the construction of
anti-selfdual Yang-Mills equations, and the application of twistor techniques
to the study of perturbative scattering amplitudes in $\mathcal{N}=4$ gauge
theory. Central to both applications is the link between 4D (self-dual) gauge
theory and holomorphic Chern-Simons (hCS) theory on twistor space. Following
Witten, we will consider the hCS theory as the field theory of the physical
modes of the B-model topological open string theory, attached to suitable
twistor space-filling D-branes.

Supertwistor space $\mathbb{CP}^{3|4}$ is a Calabi-Yau supermanifold, making
it possible to introduce a topological B-model. The corresponding open string
subsector defines a decoupled theory. The physical degree of freedom is a
$(0,1)$ component ${\mathcal{A}}$ of a gauge connection on $\mathbb{CP}^{3|4}%
$. The action is given by the holomorphic Chern-Simons form%
\bea
S_{hCS}\is \int_{\mathbb{CP}^{3|4}}\!\!\Omega\wedge\text{tr}\Bigl(\mathcal{A}%
\overline{\partial}\mathcal{A}+\frac{2}{3}\mathcal{A}\wedge\mathcal{A}%
\wedge\mathcal{A}\Bigr)
\eea
The holomorphic $(3|4)$ form $\Omega$ for the Calabi-Yau space $\mathbb{CP}%
^{3|4}$ is given by:%
\bea
\Omega\is \frac{\varepsilon_{\alpha\beta\gamma\delta}}{4!}Z^{\alpha}dZ^{\beta
}dZ^{\gamma}dZ^{\delta}\;\frac{\varepsilon_{ijkl}}{4!}d\psi^{i}d\psi^{j}%
d\psi^{k}d\psi^{l}.
\eea
where the $Z^{\alpha}$ and $\psi^{i}$ for $\alpha=1,..,4$ and $i=1,...,4$ are
respectively bosonic and fermionic homogeneous coordinates for $\mathbb{CP}%
^{3|4}$. The measure is invariant under $PSL(4|4)$.\footnote{To be slightly
more precise, the $d\psi$ entering here is more appropriately viewed as a
measure on the Berezinian in the fermionic directions \cite{Witten:2003nn}.}

The critical points of $S_{hCS}$ correspond to gauge bundles with vanishing
$(0,2)$ curvature:
\bea
\label{holo}
{\cal F}^{(0,2)}  = \overline\partial {\cal A}\!  &\! \! + \! \! & \! {\cal A} \wedge {\cal A} = 0.
\eea
In other words, they are holomorphic bundles whose complex structure is
compatible with the complex structure of the ambient $\mathbb{CP}^{3|4}$. Via
the Ward correspondence \cite{Ward:1977ta} (see also \cite{Atiyah:1977pw, Witten:1978xx}),
there is a special class of such holomorphic bundles which describe
anti-self-dual instantons on $S^{4}$. The basic equivalence is between the
following two types of mathematical objects:
\begin{align}
&  \bullet\text{ Anti-self dual Yang-Mills connections of }GL(n,%
\mathbb{C}
)\text{, defined on the four sphere } S^{4}\nonumber\\
&  \bullet\text{ Holomorphic rank }n\text{ vector bundles on }\mathbb{CP}%
^{3}\text{, that are trivial on every twistor line}\nonumber
\end{align}
The instanton number of the gauge bundle on $S^{4}$ corresponds to the second
Chern class of the vector bundle on $\mathbb{CP}^{3|4}$. Ward's construction
represents the analog for gauge fields of Penrose's `non-linear graviton'
construction of the anti- selfdual Einstein equations \cite{Penrose:1976jq}.
For more details of the Ward correspondence, we refer to the standard
literature \cite{Ward:1977ta, Atiyah:1977pw}.

The connection between the variables of twistor space, and super Yang-Mills
theory can also be seen through the expansion of $\mathcal{A}$ $\ $in the
fermionic coordinates:%
\bea
\mathcal{A}\is a+\psi^{i}\chi_{i}+\frac{1}{2!}\psi^{i}\psi^{j}\phi_{ij}+\frac
{1}{3!}\psi^{i}\psi^{j}\psi^{i}\psi^{k}\eta_{\lbrack ijk]}+\frac{1}{4!}%
\psi^{i}\psi^{j}\psi^{k}\psi^{l}b_{[ijkl]}%
\eea
The variable $\psi$ has homogeneity $+1$. From this we conclude that $a$ is a
degree zero $(0,1)$ form on bosonic twistor space, and $b$ is a degree $-4$
$(0,1)$ form. Via the correspondence between homogeneity and space-time
helicity (reviewed in footnote 11), these components of the expansion are to
be identified with the components of the $\mathcal{N}=4$ vector multiplet.

Expanding out the components and performing the fermionic integrations, the
holomorphic Chern-Simons action reduces to the the following action defined
over the bosonic twistor space $\mathbb{CP}^{3}$:%
\bea
S_{\mathrm{hCS}}\is \int_{\mathbb{CP}^{3}}\Omega^{\prime}\wedge\text{tr}%
\Bigl(b\wedge\bigl(\overline{\mathbb{\partial}}a+a\wedge a\bigr)+\phi
^{ij}\!\wedge(\overline{\partial}+a)\phi_{ij}\Bigr)+\text{fermionic terms.}%
\eea
The bosonic holomorphic three-form $\Omega^{\prime}$ is an element of
$\Omega^{(3,0)}(\mathcal{O}\left(  4\right)  )$, while the integrand is an
element of $\Omega^{(0,3)}(\mathcal{O}\left(  -4\right)  )$. The equation of
motion for $b$ selects a choice of $a$ such that the $(0,2)$ component of the
corresponding field strength vanishes. This defines a holomorphic bundle over
$\mathbb{CP}^{3}$, which includes all anti-self-dual instantons on $S^{4}$.
Hence the hCS theory on ${\mathbb{C}}\mathbb{P}^{3|4}$ is physically
equivalent to the anti-self-dual sector of $\mathcal{N} = 4$ gauge theory
\cite{Witten:2003nn}, provided that ${\mathcal{A}}$ is restricted to be flat
over each twistor line.

In the influential paper \cite{Witten:2003nn}, Witten proposed to extend this
relationship to the complete $\mathcal{N}=4$ gauge theory. The construction,
motivated by Nair's observation that MHV amplitudes are localized on twistor
lines, involved supplementing the hCS theory with D1-instantons, wrapping
holomorphic curves inside $\mathbb{CP}^{3|4}$. In this paper, we will follow a
different procedure: we will consider the pure holomorphic Chern-Simons
theory, without adding any additional degrees of freedom. Rather than
introducing D1-instantons by hand, we will study the perturbative dynamics
around a non-trivial instanton background for the gauge field $\mathcal{A}$ on
${\mathbb{CP}}^{3|4}$, with a very large instanton number. The effect of this
flux background is quite similar to adding D1-branes\footnote{Indeed, lower
dimensional branes embedded inside of higher dimensional branes dissolve into
topologically non-trivial flux configurations of the higher dimensional gauge
field.}, in that it leads to additional defect degrees of freedom that
localize on the twistor lines. But there are important differences with
Witten's proposal. Most notably, our setup introduces a UV scale, via the
size of the constituent instantons that comprise the total flux background.

Two additional comments are in order. First, note that not all holomorphic
vector bundles on $\mathbb{CP}^{3|4}$ correspond to instantons on $S^{4}$: the
critical points of the hCS\ theory also include vector bundles which do not
trivialize over twistor lines. Mathematically, such bundles leads to
additional data beyond that specified in the usual ADHM\ construction of
instantons on $S^{4}$.\footnote{See for example \cite{InstantonExtension} for
some discussion of this extension in the math literature.} Secondly, the
reduction from a string field theory to the topological B-model may also allow
for certain deformations that can not be captured purely in terms of purely
holomorphic data. In particular, like in type II theory, one can choose to
turn on a 2-form $B$-field background, which deforms the string worldsheet
action. The 2-form field acts like a $U(1)$ magnetic field, under which the
two open string end points are oppositely charged. Upon taking the decoupling
limit, each end point is forced into lowest Landau level orbits. From the
point of view of the hCS theory, turning on this 2-form $B$-field implements a
non-commutative deformation of twistor space.

\subsection{Twistor Quantization}

Twistor space and its dual space naturally combine into a quantized phase
space \cite{Penrose:1968me}. Indeed, the Heisenberg commutation relation
between positions and momenta implies that the twistors $Z^{\alpha}=\left(
\omega^{\dot{a}},\pi_{a}\right)  $ and dual twistors $\widetilde{Z}_{\beta
}=(\widetilde{\pi}_{\dot{b}},\widetilde{\omega}^{\dot{b}})$ satisfy the
commutation algebra:%
\bea
\lbrack Z^{\alpha},\widetilde{Z}_{\beta}]=\hslash\delta_{\beta}^{\alpha
}\ \ \Rightarrow\ \left\{  \
\begin{array}
[c]{l}%
\lbrack\omega^{\dot{a}},\widetilde{\pi}_{\dot{b}}]=\hslash\delta_{\dot{b}%
}^{\dot{a}}\\[2.5mm]%
\lbrack\pi_{a},\widetilde{\omega}^{b}]=\hslash\delta_{a}^{b}%
\end{array}
\right.  \label{zcom}%
\eea
The Hilbert space realization of this commutator algebra provides a
representation of $sl(4,\mathbb{C})$, the algebra of the complexified
conformal group in four dimensions, generated by respectively the
translations, conformal boosts, Lorentz rotations and dilatation%
\bea
\qquad \textstyle P_{\dot{a}a}  \is \widetilde{\pi}_{\dot{a}}\pi_{a}\ \ \ ,\ \ \ K_{a\dot
{a}}=\widetilde{\omega}_{a}\omega_{\dot{a}}\,,\nonumber\\[-2.5mm]
& & \hspace{5cm} D   ={\frac{1}{2}}\bigl(\widetilde{\pi}_{\dot{a}}\omega^{\dot{a}}%
-\widetilde{\omega}^{a}\pi_{a}\bigr).\\[-2.5mm]
J_{\dot{a}\dot{b}}  \is \widetilde{\pi}_{(\dot{a}}\omega_{\dot{b})}\ \ \, ,\  \   \text{
}\widetilde{J}_{ab}=\widetilde{\omega}_{(a}\pi_{b)}\,,\nonumber
\eea

The passage to a real space-time signature is achieved by a convention for
hermitian conjugation of the $Z^{\alpha}$ variables and the corresponding
$\widetilde{Z}_{\beta}$ variable, or equivalently, a choice of realization of
$sl(4,\mathbb{C})$. The three signatures $(++--)$, $(-+++)$ and $(++++)$
correspond to the realizations of $sl(4,%
\mathbb{C}
)$ respectively by $sl(4,%
\mathbb{R}
)$, $su(2,2)$ and $su(4)$. Hermitian conjugation in the first two cases acts
via%
\begin{align}
su(2,2)  &  :\ \ \ \ \,\omega_{\dot{a}}^{\dag}=\widetilde{\omega}%
_{a}\text{,\thinspace\ }\pi_{a}^{\dag}=\widetilde{\pi}_{\dot{a}}%
\text{\ \ \ \ and\ \thinspace\ }\hslash\in%
\mathbb{R}%
\label{su22def}\\[2mm]
sl(4,%
\mathbb{R}
)  &  :\,\left\{
\begin{array}
[c]{c}%
\omega_{\dot{a}}^{\dag}=\omega_{\dot{a}}\text{, }\pi_{a}^{\dag}=\pi_{a}\\[1mm]%
\widetilde{\omega}_{a}^{\dag}=\widetilde{\omega}_{a}\text{, }\widetilde{\pi
}_{a}^{\dag}=\widetilde{\pi}_{a}%
\end{array}
\right\}  \text{ and \ }\hslash\in i%
\mathbb{R}%
\end{align}
The reality condition on $\hslash$ is enforced by considering how hermitian
conjugation acts on the canonical commutator relations.\footnote{The above reality
conventions may be deformed by including relative complex phases. This
provides a means to analytically continue from one space-time signature to another.}
For euclidean signature, hermitian conjugation relates the $\pi$ and $\omega$ spinors via
\bea
su(4):\ \ \;\omega_{\dot{a}}^{\dag}\is \hspace{1pt}\widetilde{\pi}_{\dot{a}%
}\text{, }\ \ \pi^{\dag a}=\widetilde{\omega}^{a}\text{\ \ \thinspace
\ and\ \thinspace\ }\hslash\in%
\mathbb{R}
. \label{su4def}%
\eea
The resulting commutation relations are:%
\bea
\bigl[\spc  \omega^{\dot{a}},\omega_{\dot{b}}^{\dag}\spc \bigr]  =\hslash\delta
_{\dot{b}}^{\dot{a}} & ; & \left[  \pi_{a},\pi^{\dag b}\right]
=\hslash\delta_{a}^{b}.
\eea
Note that daggering an oscillator changes an upper index to a lower one and
visa versa. Hermitian conjugation (\ref{su4def}) acts on the conformal
symmetry generators as:%
\bea
\left(  P_{\dot{a}a}\right)  ^{\dag}=K^{\dot{a}a}\text{, }\left(  K_{\dot{a}%
a}\right)  ^{\dag}=P^{\dot{a}a}\text{, } (  J_{ab})  ^{\dag}%
=J^{ab}\text{, } (  \widetilde{J}_{\dot{a}\dot{b}})  ^{\dag
}=\widetilde{J}^{\dot{a}\dot{b}}\text{, }D^{\dag}=-D
\eea
The Euclidean theory is therefore most naturally viewed as a radially
quantized theory on an $S^{4}$. Working with respect to this realization, note
that in $su(4)\simeq so(6)$ the infinity bitwistor (\ref{infinitytwistor}) transforms as a vector
under $so(6)$. The generators which leave it invariant span a $usp(4) \simeq
so(5)$ subalgebra, which we identify with the symmetries of the four sphere. The Hermitian
$so(5)$ generators corresponding to motion along a great circle are:
\begin{equation}
\mathcal{P}_{\dot a a} = P_{\dot a a} + K^{\dot a a}.
\end{equation}

\begin{figure}[ptb]
\begin{center}
\includegraphics[
height=2.3333in,
width=2.7614in
]{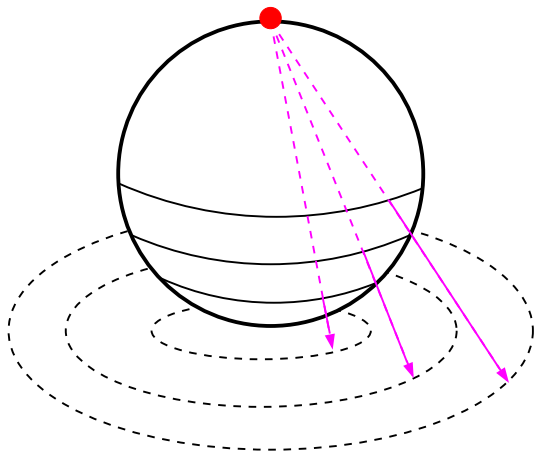}
\end{center}
\caption{Depiction of the flat space limit, corresponding to a stereographic
projection of the $S^{4}$ to $\mathbb{R}^{4}$. In the oscillator algebra this
corresponds to a Wigner-In\"{o}n\"{u} contraction of $so(5)$ to the Euclidean
Poincar\'{e} algebra.}%
\label{rescaler}%
\end{figure}

Starting from $S^4$, we can take the flat space limit by sending the radius of the sphere to infinity
and zooming in on a local flat region of the sphere.  This procedure is equivalent to performing a
Wigner-In\"{o}n\"{u} contraction, and amounts to rescaling the dual
oscillators as:
\bea
\omega_{\dot{a}}^{\dag}=\gamma^{-1}\widetilde{\pi}_{\dot{a}},\,\,\pi_{a}%
^{\dag}=\gamma\widetilde{\omega}_{a}%
\eea
where the parameter $\gamma$, the same one that appears in the $S^{4}$ bitwistor (\ref{infinitytwistor}), is being sent to zero.
In this limit,
the $so(4)$ subalgebra spanned by $J$ and $\widetilde{J}$ stays fixed, while
the Hermitian translation generators become $P+\gamma^{2}K$. The contraction
to the Euclidean Poincar{\'{e}} algebra then proceeds via $\gamma\rightarrow
0$. See figure \ref{rescaler} for a depiction of the flat space limit.

The extension to space-time supersymmetric theories is achieved by including
four fermionic oscillators $\psi_{i}$ with $\bigl\{  \psi^{i},\widetilde{\psi
}_{j}\bigr\}  =\hslash\delta_{j}^{i}$, which then provides a representation
of $psl(4|4)$. In what follows we shall often leave implicit the extension to
the fermionic case. The supersymmetric hermitian charges are the symmetries of
the supersphere $S^{4|8}$.

Finally, we emphasize that the above discussion follows the standard viewpoint
of twistor quantization \cite{Penrose:1968me}. In particular, up to
this point, the introduction of quantized twistor space does not lead to any
\textit{a priori} space-time non-commutativity. A key feature of twistor
theory is that space time is a derived notion, obtained via the correspondence
at the level of holomorphic geometry. Since (for any of the three possible
space-time signatures) the holomorphic twistor coordinates $Z^{\alpha}%
=(\omega^{\dot{a}},\pi_{a})$ all commute among each other, the notions of
twistor lines and space-time points are essentially unaffected by the quantum
relation with the dual twistors. Hence in most applications of twistor methods
-- for example, the construction of instantons and the study of scattering
amplitudes -- one does not need take into account the quantum nature of the
twistor coordinates.

\section{Instantons and Fuzzy Twistors \label{sec:FUZZ}}

In the following sections, we will introduce a twistor matrix model, which in
a large $N$ scaling limit reproduces the 4D\ continuum theory. The matrix
model is to be viewed as the low energy limit of a more UV\ complete theory
given by ordinary holomorphic Chern-Simons theory on ordinary supertwistor
space. \ By expanding around a flux background which retains the symmetries of
the 4D space-time, we obtain a low energy effective theory. Motivated, in part, by the
correspondence with the 4D Quantum Hall effect studied  in
\cite{Zhang:2001xs}, the gauge field that we will turn on is the
twistor lift of the Yang monopole \cite{Yang:1977qv, Yang:1978td}.\footnote{See \cite{SparlingQHE} for a
discussion of the relation between the 4D QHE and twistor geometry,
\cite{Fabinger:2002bk} for an embedding of the 4D QHE in string theory and \cite{Karabali:2002im}
for a discussion of the connection between the QHE effect on $S^{4}$ and $\mathbb{CP}^{3}$. For the relation between
the lowest Landau level and twistor geometry, see also \cite{Hasebe:2009vs}.}

The Yang monopole is a non-abelian generalization of the Dirac monopole to an
$SU(2)$ gauge theory defined in five dimensions \cite{Yang:1977qv, Yang:1978td}. The monopole
sits at the origin of a 5D space, surrounded by a four sphere $S^{4}$. On the $S^{4}$,
this leads to an $so(5)$ homogeneous instanton density. In addition to the
parameters specifying an ordinary instanton, the configuration space of the
Yang monopole contains an $S^{2}$. This $S^{2}$ is the space of possible
identifications between the gauge theory $SU(2)$ and a space-time $SU(2)\subset
SO(5)$ of the isometries of the $S^{4}$. A particle charged under this
background flux experiences the local geometry $S^{4}\times S^{2}$ geometry.
Globally, this $S^{2}$ fibers non-trivially over the $S^{4}$, so that for a
single instanton, we obtain a three dimensional complex projective space $\mathbb{CP}^{3}$. This
construction generalizes to $SU(n)$ gauge theory via the principal embedding
of the spin $N/2$ representation of $SU(2)$ in the fundamental of $SU(n)$
where $n=N+1$. Via the Ward correspondence, the twistor lift of the rank $n$
bundle on $S^{4}$, combined with the abelian flux, realizes a $U(1)\times
SU(n)$ gauge bundle $\mathcal{A}_{Y}$ on $\mathbb{CP}^{3|4}$.

Proceeding in the other direction, we can start form the hCS\ theory with
gauge group $G=U(nN_{c})$. Turning on a $U(1)\times SU(n)$ gauge field
${\mathcal{A}}_{Y}$ then breaks the gauge group $G$ to $U(N_{c})$. The low
energy limit is a $U(N_{c})$ hCS theory, defined on fuzzy twistor space,
coupled to fundamental matter localized on the twistor lines. From the open
string perspective, the non-commutativity arises because the open string end
points are charged under the background flux. The presence of the flux
introduces an explicit choice of scale, and the open string spectrum is gapped
with excitations of energy inversely proportional to this UV length scale \cite{TMM}. The
gapped excitations decouple at large $N$ and in the low energy limit, the
system is forced to lie in the lowest Landau level.

Our aim in this section will be to study the effective geometry experience by
particles moving in the lowest Landau level. We begin by reviewing the
topology of the Yang monopole configuration.

\subsection{Yang Monopole}

The Yang monopole is a 5 dimensional non-abelian generalization of the
Dirac monopole. The basic example consists of an abelian flux extended over
the full twistor space $\mathbb{CP}^{3}$, correlated with a homogeneous
$SU(2)$ instanton configuration along the four sphere $S^{4}$. The bundle data
is equivalently characterized in terms of the two Hopf fibrations
$S^{7}\rightarrow\mathbb{CP}^{3}$ and $S^{7}\rightarrow S^{4}$ with
fibers~$S^{1}=U(1)$ and $S^{3}=SU(2)$, respectively:
\be%
\begin{matrix}
& U(1)\,\rightarrow\,S^{7} & \cr & \mskip72mu\downarrow
\lower0.5ex\hbox{$$}\\[0.3mm]
& \mskip62mu\mathbb{CP}^{3}\cr
\end{matrix}
\qquad\  \quad%
\begin{matrix}
& SU(2)\,\rightarrow\,S^{7} & \cr & \mskip85mu\downarrow
\lower0.5ex\hbox{$$}\\[0.5mm]
& \mskip75muS^{4}\cr
\end{matrix}
\,. \label{hopfib}%
\ee
The basic Yang monopole configuration is given by the original Hopf fibration,
which carries a single unit of flux. In this case, the abelian part of the
monopole is given by the line bundle $O(1) \rightarrow\mathbb{CP}^{3}$, while the non-abelian part is given by a
single $SO(5)$ homogeneous instanton of $SU(2)$ gauge theory. We refer to the
fibrations with fiber $S^{1}$ and $S^{3}$ as respectively the abelian and
non-abelian `parts' of the Yang monopole.

A concise way to present both contributions is given by working in terms of
quaternionic coordinates. Given complex coordinates $Z^{\alpha}=\mathrm{Re}%
(Z_{\alpha})+\mathbf{i}\,\mathrm{Im}(Z_{\alpha})$ for $\mathbb{C}^{4}$, the
quaternionic plane $\mathbb{H}^{2}$ can be parameterized as $Q_{1}=Z^{1}%
+Z^{2}\mathbf{j}$ and $Q_{2}=Z^{3}+Z^{4}\mathbf{j}$, where $\mathbf{i}$,
$\mathbf{j}$ and $\mathbf{k}$ denote the usual quaternionic generators.
Modding out by $%
\mathbb{C}
^{\ast}$ on $\mathbb{C}^{4}$ realizes complex projective three-space
$\mathbb{CP}^{3}$, while modding out by the left action of the non-zero
quaternions on $\mathbb{H}^{2}$ realizes the quaternionic projective line
$\mathbb{HP}^{1}\simeq S^{4}$. In terms of the $Q$ coordinates, the abelian
and non-abelian parts of the Yang monopole can now be written as
\cite{Demler:1998pm}:%
\bea
A_{Y}=\frac{1}{2}\left(  dQ_{a}^{\dag}Q_{a}-Q_{a}^{\dag}dQ_{a}\right)  .
\label{YangMono}%
\eea
This represents a $U(1)\times SU(2)$ gauge connection via the identification
of quaternionic generators with Lie algebra elements of $SU(2)$. More general
choices of flux can be accommodated via the principal embedding of
$SU(2)\rightarrow SU(n)$, defined by identifying the spin $N/2=(n-1)/2$
representation of $SU(2)$ with the fundamental representation of $SU(n)$:
\bea
1\rightarrow N\mathbf{1}_{n\times n}\text{, }i\rightarrow-2i\mathbf{I}%
_{1}\text{, }j\rightarrow-2i\mathbf{I}_{2}\text{, }k\rightarrow-2i\mathbf{I}%
_{3},
\eea
where $\mathbf{I}_{k}$ denote the spin $N/2$ representation of the $SU(2)$
algebra. Here and in the following, the integers $N$ and $n$ are related via $
N=n+1\,.$
The abelian contribution to the Yang monopole defines a $U(1)$ gauge
connection
\bea
\mathcal{A}_{U(1)}=\frac{N}{2}(dZ_{\alpha}^{\dag}Z^{\alpha}-Z_{\alpha}^{\dag
}dZ^{\alpha}), \label{yangabel}%
\eea
with curvature equal to $N$ times the standard K\"{a}hler form on
$\mathbb{CP}^{3}$. The non-abelian contribution defines an $SU(n)$ gauge
connection over $\mathbb{HP}^{1}\simeq S^{4}$, which via the Ward
correspondence lifts to an $so(5)$ symmetric rank $n$ holomorphic vector
bundle over $\mathbb{CP}^3$.

In the following, the Yang monopole will refer to a
gauge connection on twistor space, given by the sum of the abelian flux
(\ref{yangabel}) and the twistor lift of the $SU(n)$ connection. The instanton
number of the resulting $SU(n)$ gauge field is maximized by choosing the
irreducible representation, giving an $so(5)$ homogeneous instanton with
instanton number:%
\bea
k_{\mathrm{inst}}=\frac{n(n^{2}-1)}{6}.
\eea

A somewhat more explicit description of the Yang monopole, which also brings
out its natural relation to the twistor geometry of the four sphere, is as
follows \cite{YangHolonomy}. We can parametrize the seven sphere $S^{7}$ by
means of a four-component complex vector $Z^{\alpha}$ (which we can view as a
spinor of $so(5)$), subject to the constraint
\bea
\label{ccondition}
{Z}_{\alpha}^{\dagger}Z^{\alpha}=\hslash N,
\eea
where for later convenience, we chose the radius of the $S^{7}$ equal to
$\sqrt{\hbar N}$. For now, however, $Z$ and $Z^{\dagger}$ are just classical
coordinates on $S^{7}$. The $U(1)$ action in the first Hopf fibration
$S^{7}\rightarrow\mathbb{CP}^{3}$ in (\ref{hopfib}) represents the phase
rotation
\bea
(Z,{Z}^{\dagger})\rightarrow(e^{i\phi}Z,e^{-i\phi}{Z}^{\dagger}).
\eea
Similarly, we can parametrize the four sphere $S^{4}$ with the help of a
five-component real vector $y_{A}$ (which we can view
as a vector of $so(5)$), satisfying
\bea
y^{A}y_{A}=\ell^{2}.
\eea

\def\onet{\scriptsize{1}}
\def\een{{\tiny{1}}}
\def\twee{{\tiny{2}}}
\def\drie{{\tiny{3}}}

We would like to find a parametrization of $Z_{\alpha}$ in terms of the
geometric data of the second Hopf fibration $S^{7}\rightarrow S^{4}$ in
(\ref{hopfib}). In fact, we have already found this parametrization: it is
given by the twistor correspondence for $S^{4}$ summarized in section 2.1.
Concretely, we can package the $S^{4}$ coordinates
$y^{A}$ into a $4\times4$ matrix
$\tilde\xX_{\alpha}^{\;\beta}$ as in (\ref{Xhatted}), and impose
that $\tilde\xX_{\alpha\beta}$ and $Z$ satisfy the twistor line
equation (\ref{projequation}). Indeed, in $so(5)$ spinor notation (and setting
$\ell\!=\!1$), the parametrization (\ref{xparam}) amounts to setting
$\tilde\xX=\frac{1}{2}(1-\Gamma_{A}%
y^{A})$, with $\Gamma_{A}$ the $so(5)$ gamma
matrices. The twistor line equation (\ref{projequation}) then takes the
standard form of the Hopf equation: $Z+y^{A}%
\Gamma_{A}Z=0$. This equation is explicitly solved, up to an overall phase,
via%
\bea
\label{long}
Z^{\alpha}= \frac{1}{\sqrt{\ell(\ell + y_5})} \left(  \!\!%
\begin{array}
[c]{c}%
-i y_\mu\sigma^{\mu}
\left(\!\!
\begin{array}
[c]{c}u_{1}\\
u_{2}\end{array}
\!\!\right)\\[5mm]%
(\ell+ y_5 )\left(\!\!
\begin{array}
[c]{c}u_{1}\\
u_{2}\end{array}
\!\!\right)
\end{array}
\!\!\right)  & ; &\! \left(  \!\!%
\begin{array}
[c]{c}%
u_{1}\\
u_{2}%
\end{array}
\!\!\right)  =\sqrt{\frac{\hslash N}{1+n_3}}\left(  \!\!%
\begin{array}
[c]{c}1+n_{3}\\[1mm]%
{n_1+in_2}
\end{array}
\!\!\right)
\eea
where the $n_{i}$ are normalized unit vectors on $\mathbb{R}^{3}$, defining a
unit two-sphere $n_{i}n^{i}=1$. They parametrize the twistor line above the
given point $y^{A}$ on the four sphere. Combined, the
$y^{A}$ and $n_{i}$ represent a complete coordinate
system on $\mathbb{CP}^{3}$.

Following \cite{YangHolonomy}, we may now characterize the abelian part of the
Yang monopole (\ref{yangabel}) as the Berry connection of the spinor $Z^{\alpha}$ defined over
$\mathbb{CP}^{3}$. Since the associated holonomy is just a phase, the Berry
phase Lagrangian for the spinor coordinates is
\bea
\label{berry}
\frac{1}{\hslash} L  \is \frac{i}\hslash{Z}^{\dagger}_{\alpha}\dot Z^{\alpha
}=- N \left(  \hspace{1pt} \frac{\varepsilon_{3ij}n_{i}\dot{n}_{j}}{1+n_{3}}\;
+\; \frac{{\eta_{\mu\nu}^{i} n_{i}}\hspace{1pt} y^{\mu}\dot{y}^{\nu}}{
\ell(\ell+y_{5})}\hspace{1pt} \right)
\eea
where $\eta_{\mu\nu}^{i}=\epsilon_{i\mu\nu4} + \delta_{i\mu}\delta_{\nu4} -
\delta_{i\nu}\delta_{\mu4}$ denotes the 't Hooft symbol.
Using this, we can reconstruct a gauge field flux on both the $S^{4}$
direction, and along the $S^{2}$ fiber:%
\bea
\label{yangmonop}A_{\mu} = \hspace{1pt} N\hspace{1pt} \frac{\eta_{\mu\nu}%
^{i}n_{i}\hspace{1pt} y^{\nu}}{\ell(\ell+y_{5})}\text{, }\qquad A_{5}=0,\qquad
A_{i} = \hspace{1pt} N\hspace{1pt}\hspace{1pt} \frac{\varepsilon_{3ij}n_{i}%
}{1+n_{3}}%
\eea
which defines the abelian Yang monopole connection on $\mathbb{CP}^{3}$.

Both terms in (\ref{yangmonop}) have a familiar form. The $S^{2}$ part $A_{i}$
represents the constant magnetic flux through a unit two sphere, produced by a
Dirac monopole of charge $N$ at its center. It arises because the two Hopf
fibrations in eqn (\ref{hopfib}) are related to each other via the basic Hopf
fibration $S^{3}\rightarrow S^{2}$ that embeds $U(1)$ inside $SU(2)$.
Similarly, the $S^{4}$ part $A_{\mu}$ looks like the abelianized twistor lift
of the basic $SU(2)$ instanton -- or rather its embedding via the spin $N/2$
representation inside $SU(n)$ -- via the identification
\bea
n_{i}\;\leftrightarrow\;-\frac{2i}{N}\hspace{1pt}\mathbf{I}_{i}. \label{nmap}%
\eea
Indeed, making this substitution inside of the $S^{4}$ part $A_{\mu}$ in
(\ref{yangmonop}) gives a maximal homogenous $SU(n)$ instanton on $S^{4}$.
Again we see that the two components of the monopole, the
abelian flux (\ref{yangabel}) and the non-abelian homogeneous instanton background on $S^4$,
are directly linked via the substitution (\ref{nmap}) of the $S^2$ coordinate with an
$SU(2)$ generator embedded in $SU(n)$.
The complete rank $n$ Yang monopole $\mathcal{A}_{Y}$ is the sum of the abelian
magnetic field (\ref{yangmonop}) with the twistor lift of the homogeneous
$SU(n)$ instanton.

\subsection{Lowest Landau Level}

Let us look a little more closely at the effective geometry experienced
charged particles in the abelian flux background. At low energy, the particles are forced into their
lowest Landau level (LLL). This system was analyzed in detail in
\cite{Zhang:2001xs, YangHolonomy}.  The LLL states are most  succinctly characterized as the Hilbert
state obtained by canonical quantization of the abelian Berry holonomy Lagrangian (\ref{berry}).

We can do this in two different ways. We can first quantize the $Z_\alpha$ coordinates, and then impose
the constraint (\ref{ccondition}) or we can first solve the constraint, as done via the $S^4 \times S^2$
parametrization (\ref{long}). Let us first follow the second route.
From the second form of the Berry action in (\ref{berry}),
we can read off the commutation relations among the $n_i$ and $y_\mu$, by following the
usual rules of canonical quantization. One finds coordinates which satisfy the following
commutator algebra\cite{Zhang:2001xs, YangHolonomy}
\bea
\label{comalgebra}
\left[  n_{i},n_{j}\right]  =\frac{2}{N}\hspace{1pt}\epsilon_{ijk}%
n_{k}\,,\text{ \ \ }\left[  y_{\mu},y_{\nu}\right]  =\frac{\ell^{2}}%
{2N}\hspace{1pt}\eta_{\mu\nu}^{i}n_{i},\text{ \ \ }\left[  n_{i},y_{\mu
}\right]  =\frac{2}{N}\hspace{1pt}\eta_{i\mu\nu}y^{\nu}.
\eea
Rescaling the $n_{i}$ coordinates that paramterize the two sphere, we see that they turn into  generators of an $su(2)$ algebra that rotates the $S^2$.
Since the rescaled $S^2$ has radius $N$, we learn that the $su(2)$ acts via a spin $N/2$ representation.
The $S^2$ has turned into a non-commutative sphere with $n = N+1$ fuzzy points. Moreover, we see that the $su(2)$ algebra does
not commute with the position coordinates $y^\mu$, but instead act like the generators of chiral space-time rotations.
The space-time coordinates $y_{\mu}$  coordinates also do not commute among each other:
their commutator is a generator of a chiral $su(2)$ rotation. We will refer to the non-abelian
commutator algebra (\ref{comalgebra})  later on, when we begin our study of
the low energy physics of the hCS theory with flux.

The above discussion also clarifies the mapping (\ref{nmap}) between the abelian and non-abelian flux: it identifies the spin $N/2$
representation of chiral $su(2)$ rotations with $su(2)$ gauge generators embedded inside an $SU(n)$ gauge group. The replacement (\ref{nmap})
results in a homogeneous, $so(5)$ invariant instanton configuration, by virtue of the fact that we can combine the chiral
$su(2)$ space-time rotations with global $su(2)$ rotations, that acts on the Lie algebra labels of the $SU(n)$ gauge field.
Indeed one can show that the total space spanned by the Landau level wave functions preserves an
overall $so(5)$ symmetry, and that moreover, the lowest Landau level transforms as a representation of $su(4)$ \cite{Zhang:2001xs}.

We now turn to a more practical description of the LLL states: we first quantize the $Z_\alpha$ coordinates,
and then afterwards impose the constraint (\ref{ccondition}).  This procedure identifies the
LLL states with the space of Planck cells, or fuzzy points, on non-commutative $\mathbb{CP}^3$.
The $su(4)$ symmetry of the LLL level is manifest in this description.\footnote{The approach to non-commutative geometry we consider has been
developed in \cite{FUZZ}, and we refer the interested reader there for
additional details (see also \cite{Iqbal:2003ds, Saemann:2006gf}).}

From the first form of the Berry Lagrangian (\ref{berry}) we immediately read off that the four coordinates $Z^{\alpha}$ act
as bosonic oscillators with commutators:%
\bea
\label{zcom}
\bigl[  Z^{\alpha},Z_{\beta}^{\dag}\bigr]  \is \hslash\spc \delta_{\beta}^{\alpha}.%
\eea
These generate a Fock space obtained by acting with the $Z_\beta^\dag$ oscillators on the vacuum state $\left\vert 0\right\rangle $ annihilated by the $Z^{\alpha}$ oscillators.
Each basis state in the Fock space represents one Planck
cell of the non-commutative
space $\mathbb{C}^{4}$, Since this space is non-compact, the associated Hilbert space  $\mathcal{H}_{\mathbb{C}^{4}}$ is infinite dimensional.

Now let us impose the constraint (\ref{ccondition}). At the quantum level, this will automatically lead to a projection from $\mathbb{C}^4$ onto the
complex projective projective space $\mathbb{CP}^{3}$.
Introduce the level operator $H_{0} = Z_{\alpha}^{\dag}Z^{\alpha}.$
This operator has an integer spectrum, given by the sum of the oscillator levels of the ${\cal Z^I}$ number
eigenstates.  The constraint (\ref{ccondition}) is now imposed at the level of the Hilbert states
\bea
\label{level}
 H_{0}\left\vert \Psi\right\rangle
\is \hbar N\left\vert \Psi\right\rangle  \quad ; \quad H_0 = Z^\dag_\alpha Z^\alpha\, .
\eea
We denote by $\mathcal{H}_{\mathbb{CP}^{3}%
}(N)$ the Hilbert space of states that satisfies this condition.
At a geometric level, states in $\mathcal{H}_{\mathbb{CP}^{3}}(N)$ represent holomorphic sections of the degree $N$ line
bundle $\mathcal{O}_{\mathbb{CP}^{3}}(N)$.
Note that the level constraint $H_0 = N$ indeed eliminates one complex dimension:  it
fixes the absolute value of $Z^\alpha$ but also implements the $U(1)$ invariance under phase rotations $Z^\alpha \to e^{i
\varphi} Z^\alpha$. Eqn (\ref{level})  is the non-commutative way of realizing $\mathbb{CP}^{3}$ as the K\"{a}hler
quotient $\mathbb{C}^{4}/\!/U(1)$.
Observe that the states of $\mathcal{H}_{\mathbb{CP}^{3}}(N)$ are created by
homogeneous degree $N$ polynomials in the $Z^{\dag}$'s. We can immediately count that
\bea
\dim\mathcal{H}_{\mathbb{CP}^{3}}(N) 
\is \frac{(N+1)(N+2)(N+3)}{6}\equiv k_{N}.
\eea
This formula counts the number of Planck cells that fit inside the compact space $\mathbb{CP}^{3}$, or equivalently,
the relative inverse volume of a fuzzy twistor point.
We will focus on the leading behavior in the limit of large $N$.
In this limit, the size of the Planck cells,  {\it i.e.} the scale of non-commutativity,
tends to zero relative to the total size of the projective space $\mathbb{CP}^{3|4}$.

The supersymmetric generalization of the story is straightforward. In addition to the
bosonic oscillators, introduce four fermionic oscillators $\psi^{i}$
satisfying:%
\bea
\bigl\{  \psi^{i},\psi_{j}^{\dag}\bigr\}  \is \hslash\delta_{j}^{i}\text{.}%
\eea
The Hilbert space of points for fuzzy $\mathbb{C}^{4|4}$ is given by the Fock
space of states $\mathcal{H}_{\mathbb{C}^{4|4}}$ generated by $\mathcal{Z}%
_{J}^{\dag}$. The restriction to fuzzy $\mathbb{CP}^{3|4}$ is achieved by
introducing the Hamiltonian constraint:
\bea
H_{0}\is {Z}_{\alpha}^{\dag}{Z}^{\alpha}+\psi_{i}^{\dag}\psi^{i}.
\label{superconstraint}%
\eea

\subsection{Space-Time and Locality\label{ssec:partrans}}

One of the key features of twistor theory is that space-time physics
is a derived notion, obtained via the correspondence between complex lines in
twistor space and space-time points.  Since this is a correspondence at the
level of the holomorphic geometry, it is essentially left intact by the non-commutativity.

Let us first discuss holomorphic subspaces in $\mathbb{CP}%
^{3}$. In commutative geometry, a holomorphic divisor $S\subset\mathbb{CP}%
^{3}$ is specified by the vanishing locus of a degree $d$ holomorphic
polynomial $f(Z^{\alpha})$. In the fuzzy setting, the Hilbert space of points
for $S$ is given by states of $\mathcal{H}_{\mathbb{CP}^{3}}(N)$ annihilated
by $f(Z^{\alpha})$. Let us note that generically this space is non-empty,
since $f:\mathcal{H}_{\mathbb{CP}^{3}}(N)\rightarrow\mathcal{H}%
_{\mathbb{CP}^{3}}(N-d)$ is a linear map to a vector space of smaller
dimension. Intersections of divisors proceed in a similar fashion. Given two
holomorphic polynomials $f_{1}$ and $f_{2}$ of respective degrees $d_{1}$ and
$d_{2}$, the space of states annihilated by both $f_{1}$ and $f_{2}$ defines
the Hilbert space of points for a fuzzy curve in $\mathbb{CP}^{3}$. Finally,
given polynomials $f_{1}$, $f_{2}$ and $f_{3}$, we generically obtain a
discrete collection of fuzzy points. See \cite{FUZZ} for further discussion of
intersection theory on fuzzy spaces.

To any point $p$ of \emph{commutative} twistor space, we can associate a
corresponding Hilbert space ${\cal H}_p$. To see this, note
that $p$ can be viewed as the intersection of three linear divisors,  $f_{\alpha}^{(i)}Z^{\alpha}=0$ for $i=1,2,3$.
The space of states annihilated by all three polynomials defines a
one-dimensional Hilbert space. This also provides a fiber bundle over the
commutative $\mathbb{CP}^{3}$ which we can identify with the abelian part of
the Yang monopole.

Especially important for the twistor correspondence is that for any given point $(x^{\dot{a}a},\theta^{ia})$ on the
\emph{commutative} complexified Minkowski space, there is a
corresponding fuzzy $\mathbb{CP}^{1}(x,\theta)$, with an $n=N+1$ dimensional Hilbert space ${\cal H}_{x,\theta}(N)$
spanned by all states $| x,\theta \ra$ that satisfy the annihilator equations:%
\bea
\label{line}
\left(  \omega^{\dot{a}}-ix^{\dot{a}a}\pi_{a}\right)  \left\vert
x,\theta\right\rangle =0\quad ; \quad \left(  \psi^{i}-\theta^{ia}\pi_{a}\right)
\left\vert x,\theta\right\rangle =0
\eea
These equations implement the commutative twistor line equations (\ref{bosonicline}) as a linear
projection on the Hilbert space ${\cal H}_{x,\theta}(N)$. This will be a key point of our further discussion:
although twistor space has become non-commutative, there is still a continuous moduli space of twistor lines.
Similarly we can define a Hilbert
space of dual bra states $\mathcal{H}_{x,\theta}^{\vee}$ via:%
\bea
\label{dline}
{\left\langle x,\theta\right\vert \bigl(  \omega_{\dot{a}}^{\dag}-ix_{\dot{a}%
a}\pi^{a\dag}\bigr)  =0} \quad ; \quad
{\left\langle x,\theta\right\vert \bigr(
\psi_{i}^{\dag}-\theta_{ia}\pi^{a\dag}\bigr)  =0}  \, .
\eea
The index $i$ of $\theta_{ia}$ has been lowered using the infinity twistor $\mathcal{I}_{IJ}$.
We can thus speak of a space of ket states $\mathcal{H}_{x,\theta}$
and bra states $\mathcal{H}_{x,\theta}^{\vee}$ located at any given space-time point
$(x,\theta)$.  

Given a specific twistor location $\left(\omega^{\dot a},\pi_a,\psi^i) = (x^{\dot a b} \pi_b,\pi_a, \theta^{ia}\pi_a\right)$  on the twistor line  at $(x,\theta)$,
we can associate  an element of $\mathcal{H}_{x,\theta}$ via the condition:%
\bea
\label{coher}
\left(  \pi_{2}-\lambda\pi_{1}\right)  \left\vert x,\theta;\lambda
\right\rangle =0.
\eea
Here $\lambda = \pi_2/\pi_1$ denotes the affine coordinate on the $\mathbb{CP}^1$. When combined, the conditions (\ref{line}) and (\ref{coher}) select a one dimensional
Hilbert space within ${\cal H}_{\mathbb{CP}^{3|4}}(N)$.

On the finite radius $S^{4}$, there is a closely related way to define
position states $\left\vert y,\theta;\lambda\right)$, where $y^{A} = (y^\mu, y^5)$ denote the $S^4$ coordinates
introduced in eqn (\ref{xparam}), as follows. Begin with a
state on the $\mathbb{CP}^{1|0}$ at the origin where $y_\mu = 0$ and $y_5=1$. This state $\left\vert 0,0;\lambda
\right)$ is annihilated by  $\omega^{\dot a}$, $\psi^i$ and $\pi_{2}-\lambda\pi_{1}$.
Next, we introduce a finite symmetry transformation of the supersphere $S^{4|8}$,
which we denote by $\mathcal{R}\left(  y,\theta\right)$, which maps the origin to the point $(y,\theta)$.
Its bosonic part is an $SO(5)$ rotation. We can thus obtain
general position eigenstates by applying the symmetry transformation
\bea
\label{transport}
\left\vert x,\theta;\lambda\right)  =\mathcal{R}\left(  x,\theta\right)
\left\vert 0,0;\lambda\right)  .
\eea
This defines a unitary parallel transport operation. In other words,  each point $(y,\theta)$
on the sphere $S^{4|8}$ now comes equipped with an $n$ dimensional linear space ${\cal H}_{y,\theta}$ which via (\ref{transport})
turns into an $su(n)$ bundle over  $S^{4|8}$, or more specifically, the spin $N/2$ lift of an $su(2)$ bundle over $S^{4}$. This is the
gauge bundle for the non-abelian part of the Yang monopole.

Since the Hilbert space ${\cal H}_{\mathbb{CP}^{3|4}}(N)$ is finite dimensional, it is evident that the
states associated with different space-time points can not all be independent. Rather, we should expect that
states at nearby space-time points have a non-zero overlap. The short distance scale $\ell_{pl}$,
where locality breaks down,  is determined by the overlap between two neighboring
states $\left\vert y,\theta;\lambda\right)  $ and $\left(  y^{\prime},\theta;\lambda\right\vert $.
A simple calculation, outlined in  \cite{TMM, GMM}, shows that for $y$ and $y'$ close to each other
(and close to the south pole where $y^\mu \ll y_5$)
\bea
\label{overlap}
\left(  y^{\prime},\theta,\lambda|y,\theta; \lambda\right)  =\exp\Bigl(  -\frac{N}{8\ellcc^{2}%
}\left\vert y-y^{\prime}\right\vert ^{2}\Bigr)  \times \left(  0, \theta,\lambda|0,\theta; \lambda\right).
\eea
In the large $N$ limit, the prefactor approaches a regulated delta function, smeared out over a small region on $S^{4|8}$
of linear size
\bea
\ell_{pl} \simeq \frac{\ellcc}{\sqrt{N}}. \label{lplanck}%
\eea
As suggested by our notation, this will play the role of the Planck length.

An equivalent way to see the breakdown of locality is by considering the range of allowed angular momenta
of functions on $S^{4}$. To this end, we introduce a holomorphic position operator $\hat{X}_{\alpha\beta}$ which
acts on a matrix $\Phi$ as:%
\bea
\label{xdef}
\hat{X}_{\alpha\beta}  \Phi=Z_{\alpha}\Phi Z_{\beta}^{\dag}-Z_{\beta
}\Phi Z_{\alpha}^{\dag}.
\eea
This definition is motivated by the commutative relation (\ref{twisquadX}) between the space-time coordinates $X_{\alpha\beta}$ and pairs of
twistors $Z_\alpha$ and $W_\beta$. The operator $\hat{X}_{\alpha\beta}$ indeed selects a space-time point, as follows.
It is not hard to show \cite{GMM} that the eigen operators, defined via
\bea
\label{xeigen}
\hat{X}_{\alpha\beta}\Phi(x)=x_{\alpha\beta}\hat{X}_{0}
\Phi(x)
\eea
where $\hat{X}_{0}=I^{\alpha\beta}\hat{X}_{\alpha\beta}$ and $\xX_{\alpha\beta}$ is a c-number, are annihilated from the left and right by the associated
twistor line conditions (\ref{line}) and (\ref{dline}).  This suggests that we can construct a space of functions associated with the space-time as
follows.

Introduce the hermitian conjugate coordinates $\hat{X}_{\alpha\beta}^{\ast} \Phi  = Z^\dag{\!\! }_{[\alpha} \Phi Z_{\beta]}$. Starting from the vacuum state $\left\vert
0\right\rangle \left\langle 0\right\vert $, we can repeatedly act via
$\hat{X}_{\alpha\beta}^{\ast}$. After acting 
$N$ times, we obtain a subset of operators on ${\cal H}_{\mathbb{CP}^{3|4}}(N)$, which we
denote by $\mathcal{H}_{S^{4}}(N)$. These matrices form an irreducible
representation of $su(4)$, given by a Young tableau with two rows of $N$
boxes. The dimension of this irreducible representation is:%
\bea
\dim\mathcal{H}_{S^{4}}(N)=\frac{1}{12}\left(  N+3\right)  \left(  N+2\right)
^{2}(N+1)=\frac{1}{2}k_{N}\spc (n+1)
\eea
which is the number of independent functions on $S^{4}$ up to a maximal
angular momentum. We can further decompose this into representations of
$so(5)$, which are  labelled by Young tableaux with two rows of
lengths $r_{1}$ and $r_{2}\leq r_{1}$, with dimension
(see e.g. \cite{Ramgoolam:2001zx}) $D(r_{1},r_{2})=\frac{1}{6}\left(  r_{1}+r_{2}+2\right)  \left(  r_{1}%
-r_{2}+1\right)  \left(  3+2r_{1}\right)  \left(  1+2r_{2}\right)  .$\footnote
{Note that LLL states transform in the $so(5)$ irrep with $r_{1}=r_{2}=N/2$, with dimension
$k_{N}$.}
The functions on a $S^{4}$ up to a cutoff angular momentum $N$ are given by
the direct sum of the $r$-fold symmetric product on the vector representation,
i.e. representations with $r_{2}=0$:
\bea
\mathcal{H}_{S^{4}}=\underset{r=0}{\overset{N}{\oplus}}D(r,0).
\eea
In \cite{Ramgoolam:2001zx, Castelino:1997rv} this
space of states was interpreted as the space of functions for a fuzzy $S^{4}$. Here, we are simply
considering a theory on an $S^{4}$ where we truncate the spherical harmonics
to a maximal angular momentum.
Since the space of functions scales as $N^{4}$, the number of pixels which can
be reconstructed is of order $\sqrt{N^{4}}$. The minimal length scale which
can be resolved by this angular momentum cutoff is  $\ell_{pl}$ of equation (\ref{lplanck}).

\section{Twistor Matrix Model \label{sec:TMM}}

After collecting the main geometric ingredients,  we are finally ready to introduce
the physical system of interest. In this section we wish to extract the
effective matrix model that captures the low energy dynamics of holomorphic Chern-Simons
theory in the background of the Yang monopole flux. The theory is formulated in terms of finite
size matrices, and so in particular describes a theory on space-time with a truncated number
of degrees of freedom. See \cite{Lechtenfeld:2005xi} for earlier discussion
on\ a potential matrix model characterization of twistor string theory.

\subsection{Chern-Simons with Flux}

We wish to study the perturbative dynamics of a $U(n N_c)$
holomorphic Chern-Simons theory in the Yang monopole background.
As explained, this monopole has two components: an abelian flux
given by (\ref{yangabel}), and a non-abelian flux given by the twistor lift
of a homogeneous $SU(n)$ instanton with maximal instanton charge $k_N$.
Both components have an important influence on the low energy effective
theory. We first discuss the abelian flux.

The proper way to view the abelian flux is through the lens of the topological B-model.
In any ordinary Chern-Simons gauge theory, all fields transform in the
adjoint of the gauge group and the overall $U(1)$ gauge group factor decouples. Hence
turning on an abelian flux would not have any effect on the low energy dynamics.
Instead, we will consider the holomorphic Chern-Simons theory as defined via an
open topological string theory, or equivalently, by taking the zero slope limit of the
open twistor string theory introduced and studied in \cite{Berkovits:2004hg, Dolan:2007vv}.
In the open string theory one can choose to turn on a $B$-field background, which
acts like an abelian magnetic field, under which the open string end points are
oppositely charged. In the zero slope limit, each end point is forced into a
lowest Landau level orbit, or equivalently, is compelled to occupy a state
in ${\cal H}_{\mathbb{CP}^{3|4}}$.

This gives us a first hint of what the low energy theory should
look like: by turning on the abelian flux, we have deformed the original $U(n N_c)$
holomorphic Chern-Simons theory into a non-commutative $U(n N_c)$ hCS theory defined on
twistor space $\mathbb{CP}^{3|4}$, deformed via commutation relations (\ref{zcom}).
What does this deformation correspond to in space-time?
Via the appropriate generalization of the Penrose-Atiyah-Ward correspondence,
classical solutions to the non-commutative hCS theory, that is, holomorphic $U(nN_c)$
gauge bundles with $F^{(0,2)}\! =\! 0$, correspond to
instanton backgrounds on a fuzzy $S^4$. This fuzzy four sphere is
characterized by the commutator algebra
(\ref{comalgebra}), or equivalently,
by the collection of fuzzy twistor lines introduced in the previous section.
The most symmetric  instanton background is the non-abelian
Yang monopole with rank $n$ and maximal instanton number $k_N$.

What are the low energy consequences of turning on the non-abelian flux?
A first obvious consequence is that the unbroken low energy gauge group
is reduced from $U(nN_c)$ to $U(N_c)$. So we should expect the low energy theory
to contain a non-commutative hCS sector with gauge group $U(N_c)$.
But we should also look for other low energy
remnants of the instantons. From a D-brane perspective, adding an
instanton background  amounts to adding $k_N$ co-dimension four D-branes,
one brane per constituent instanton. We will refer to the elementary instantons,
or co-dimension four D-branes, as defects. Based on our experience with
intersecting branes, we should thus anticipate the  existence of additional
defect degrees of freedom associated to the non-abelian instantons.
An alternative and possibly more direct explanation for the existence
of defect modes comes from considering the ADHM description of instantons.

One guiding principle for fixing the form of the matrix model is that the low
energy theory should be compatible with the symmetries preserved by the Yang
monopole. Since this configuration comes with a choice of scale,
specified by the instanton density, we should expect conformal symmetry to
be broken.  However, we do require the bosonic $SO(5)$ symmetry to be manifest.
Moreover, because of the close link between hCS theory and instantons on $S^{4}$,
we expect that, in an appropriate limit, the configuration space of the matrix model
should reproduce the ADHM\ construction of instantons.

This ADHM correspondence provides important guidance, so let us make the connection
a bit more precise. Since the Yang monopole breaks conformal symmetry,
the  low energy theory is aware of  the radial size $\ell$ of the $S^4$.
There is only one scale because the instanton density on $S^{4}$ is
homogeneous. However, if we give up
$SO(5)$ symmetry, we can separate the scale defined by the abelian flux from
the scale defined by the non-abelian flux.

In fuzzy twistor terms, this is done as follows. The LLL states
admit the action of an $su(4)$ algebra, generated by the hermitian charges $J$,  $\tilde{J}$,
$P+K$, $i(P-K)$ and $iD$.  The hermiticity convention $P^\dag = K$ depends on the
strength of the abelian flux. The $su(4)$ symmetry is broken to $so(5)$ via the introduction of the
infinity twistor (\ref{infinitytwistor}), provided we make the $so(5)$ invariant choice
$\gamma=1$.  As we will see, the appearance infinity twistor is linked to the presence of
the non-abelian flux.  Taking $\gamma\neq 1$  breaks $so(6)$ further down
to $so(4)$. These infinity twistors are associated with $SU(n)$ instanton backgrounds that
are homogeneous relative to a four sphere of a different radius, i.e. with non-hermitian
symmetry  generators $P+\gamma K$ relative to the inner product
set by the abelian flux. Taking the limit $\gamma \to 0$ corresponds to non-abelian
instanton configurations which are centered around the south pole of the $S^{4}$.
In this limit the abelian flux gets diluted, and we expect to make contact with the
ADHM construction of $SU(n)$ instantons. This check is performed in section \ref{sec:ADHM}.

\subsection{The Matrix Model}

Let us write the matrix model action. It consists of two terms. The first term is the non-commutative
holomorphic Chern-Simons action and the second term is the defect action
\begin{align}
\label{mmacto}
S_{\mathrm{MM}} & \, = \; S_{\mathrm{hCS}}({\cal A})+S_{\mathrm{defect}}(Q,\tilde{Q}, {\cal A})\text{.}\\[4mm]
\label{mmactt}
S_{\mathrm{hCS}} ({\cal A}) \!\!\!\!\!\!\!\!\!\! &\;\;\;\;\;\; =   
\text{Tr} 
\Bigl(
\Omega_{\alpha\beta\gamma}D^{\alpha}D^{\beta}D^{\gamma
}\Bigr)  _{\!\psi^{4}}\\[3.5mm]
\label{mmactd}
S_{\mathrm{defect}} &  (Q,\tilde{Q}, {\cal A})   =\text{Tr} 
\Bigl(\mathcal{I}_{I\!\spc J}\widetilde{Q}\spc \mathcal{D}^{I}Q {\cal Z}^{J}\Bigr)
\end{align}
where $\Omega_{\alpha \beta \gamma}$ is a three index anti-symmetric tensor which is the
non-commutative analogue of the holomorphic three-form. Here the symbol  ${\rm Tr}$ is
the trace over the Hilbert space ${\cal H}_{\mathbb{CP}^{3|4}}(N)$ tensored
with the $u(N_c)$ color space. We give the precise definition of each  matrix variable in the following
subsections. Roughly, each symbol defines a linear operator that acts on ${\cal H}_{\mathbb{CP}^{3|4}}(N) \otimes u(N_c)$.
The $D^{\alpha} = Z^\alpha + {\cal A}^\alpha$ are non-commutative versions of anti-holomorphic covariant derivatives and
$\mathcal{D}^{I}$ is an extension of this derivative to superspace. The $Q$ and $\widetilde{Q}$'s are the defect
modes, and ${\cal Z}^I = (Z^\alpha, \psi^i)$ are the non-commutative supertwistor coordinates.
Finally, ${\cal I}_{IJ}$ is the supersymmetric infinity bi-twistor (\ref{supinftwist}), and the
subscript $\bigl(...\bigr)_{\psi^4}$ indicates the projection onto the top superfield component.

All matrix variables in the action (\ref{mmacto})-(\ref{mmactd}) have direct geometric meaning as functions and sections of bundles
over twistor space. Functions correspond to maps from
$\mathcal{H}_{\mathbb{CP}^{3|4}}(N)$ to $\mathcal{H}_{\mathbb{CP}^{3|4}}(N)$. A convenient representation of such maps is in
terms of a power series in the oscillators, which we may normal order as $M=
M_{mn|ab}\hspace{1pt} (Z^{\dag})^{m}( \psi^{\dag}%
)^{a}Z^{n}\psi^{b}.$
Matrix multiplication then corresponds to multiplying successive normal
ordered power series.
Similarly, we can describe sections of degree $l$ line bundles by
rectangular matrices. These correspond to maps from $\mathcal{H}_{\mathbb{CP}^{3|4}}(N+p)$ to $\mathcal{H}_{\mathbb{CP}^{3|4}}(N+q)$, where the net degree is $l=q-p$.
The corresponding map has degree $N+p$ in the $Z$ oscillators, and $N+q$ in
the $Z^{\dag}$ oscillators. Hence, it can be identified in the commutative
geometry with the corresponding section of $O(l)$. Note, however, that there
is some freedom in how to assign $p$ and $q$ to a degree $l$ line bundle. This
is because complex conjugation and dualization of a bundle are naturally
related by Hermitian conjugation of operators. Finally, integration of a
product of such functions or sections proceeds by tracing over
the appropriate Hilbert space \cite{FUZZ}.

In the matrix action, the covariant derivatives $D^\alpha$ and ${\cal D}^I$ are (0,1) forms  on $\mathbb{CP}^{3|4}$
and like the coordinates $Z_\alpha$ and ${\cal Z}^I$ (which also act like $\overline{\partial}$ operators)
 and describe non-square matrices of degree $-1$, that change the rank from $N+p$ to $N+p-1$. The $Q$'s and $\tilde{Q}$
variables are sections of a degree 1 line bundle, and thus increase the rank from $N+p$ to $N+p+1$. Modulo such small shifts,
all symbols in (\ref{mmacto})-(\ref{mmactd}) are $k_N$ times $k_N$ matrices.

\subsection{Matrix Chern Simons \label{sec:MatCS}}

To motivate the form of the holomorphic Chern-Simons action, it is actually helpful
to at first enlarge the field content, to cover all $\mathbb{CP}^{3|4}$'s up
to level $N$. This provides an eight-dimensional formulation, as in integral over a ball of square radius $N$ inside $\mathbb{C}^{4|4}$,
 from which the hCS contribution arises as a
boundary term at level $N$. We introduce a fuzzy ball $\mathbb{B}%
^{4|4}$ with a Hilbert space of points:%
\be
\mathcal{H}_{\mathbb{B}^{(4|4)}}=\underset{M=0}{\overset{N}{\oplus}%
}\mathcal{H}_{\mathbb{CP}^{3|4}}(M)
\ee
The matrices $D^\alpha$ correspond to $(0,1)$-forms on $\mathbb{B}^{4|4}$. We focus on
the bosonic part of this differential form content. Each $D^{\alpha}$ is
specified by a power series in the $Z^{\dag}$'s, $Z$ and $\psi$ which we
schematically write as: $D^{\alpha}=D_{lm|a}^{\alpha}(Z^{\dag})^{l}Z^{m}\left(  \psi\right)  ^{a}$
with specified coefficients $D_{lm|a}^{\alpha}$.
Let us note that this expansion provides a convenient
way to deal with matrices which are of different sizes. The field content is
specified in terms of the direct sum of $K_{M}\times K_{M+1}$ matrices for
$M=0$ up to $M=N$, which we denote by:%
\be
D^{\alpha}\equiv\underset{M=0}{\overset{N}{\oplus}}D_{M\times(M+1)}^{\alpha}%
\ee
Each term in this direct sum is controlled by the same
expansion coefficients $D_{lm|a}$. The degrees of freedom of the matrix model
truncates at finite order set by $N$. The $D^{\alpha}$ transform in the
adjoint representation of the $U(N_{c})$ gauge group, and are
to be thought of as covariant derivatives.
To ensure that the gauge fields project to $(0,1)$
forms on $\mathbb{CP}^{3|4}$, they are subject to the constraint
\cite{FUZZ}:%
\begin{equation}
Z_{\alpha}^{\dag} D^{\alpha} = \hbar N \label{constrainer}.
\end{equation}
Expanding around $D^{\alpha} = Z^{\alpha} + \mathcal{A}^{\alpha}$, this becomes
the condition $Z_{\alpha}^{\dag} \mathcal{A}^{\alpha} = 0$. This projects out
the direction orthogonal to $\mathbb{CP}^{3|4}$ inside of
$\mathbb{C}^{4|4}$. See \cite{FUZZ} for further discussion.

The fuzzy holomorphic Chern-Simons action can now be defined as a trace  over the 8-dimensional ball $\mathbb{B}^{4|4}$
\bea
\label{haas}
S_{\mathrm{hCS}} ({\cal A}) \is  \frac{1}{g^{2}}
\text{Tr}_{\hspace{1pt}\strut{\mathbb{B}}^{4|4}}\!
\Bigl(
\varepsilon_{\alpha\beta\gamma\delta}D^{\alpha}D^{\beta}D^{\gamma}D^{\delta
}\Bigr)  _{\!\psi^{4}}
\eea
where we have introduced $g$, the gauge coupling of the matrix model.
This is the non-commutative version of the continuum action
$\int_{\mathbb{B}^{4|4}}d^{4}\psi d^{4}Z {\rm tr} \left(\mathcal{F}\wedge\mathcal{F}\right) $ on $\mathbb{B}^{4|4}$ where
$\mathcal{F}=\overline{\partial}\mathcal{A}+\mathcal{A}\wedge\mathcal{A}$.
This continuum action is a total derivative, and so
integrates to just a boundary term, which is the hCS\ action.\footnote{D.
Skinner has also considered an eight-dimensional formulation of hCS theory.}
A similar effect occurs in our case: the matrix operator inside the trace in (\ref{haas}) naively takes the form of a total commutator:
\be
\varepsilon_{\alpha\beta\gamma\delta}\bigl[D^{\alpha},D^{\beta}D^{\gamma}D^{\delta}\bigr]
\ee
and thus should have a vanishing trace. However, the $D^\alpha$'s are non-square matrices that lower the rank of the
Hilbert space by one unit: they map ${\cal H}_{\mathbb{CP}^{3|4}}(N+p)$ to  ${\cal H}_{\mathbb{CP}^{3|4}}(N+p-1)$.
So if we cycle $D^\alpha$ around the trace, it lowers the rank of the Hilbert space over which the trace is taken on one
unit. It is easy to see that, via this mechanism, the trace over all states inside the ball $\mathbb{B}^{4|4}$ cancel out,
but that one is in fact left with  boundary term at level $N$. This boundary term includes the hCS action as written
in (\ref{mmactt}).

There is in fact a subtlety in this partial integration argument, that will play an important role later on.
The eight-dimensional form (\ref{haas}) of the holomorphic Chern-Simons action is manifestly
invariant under the gauge transformations:%
\bea
D^\alpha \rightarrow e^{ih}{D}^{\alpha}e^{-ih}\text{.}
\label{gaugetransform}%
\eea
where $h=T^{A}h_{A}$ is a $u(N_{c})$ Lie algebra element and $h_A$ are arbitrary degree zero  polynomials in
the $Z^{\dag}$ and $Z$'s. Hence the result after the partial integration should also be gauge invariant.
However, as the attentive reader may already have noticed, the action (\ref{mmactt}) is not automatically
gauge invariant, since, even though it is a color singlet, the holomorphic three-form $\Omega_{\alpha \beta \gamma}$ does
{\it not} commute with general non-commutative gauge parameters $h_{A}(Z,Z^\dag)$. This effect is subleading in $1/N$,
and thus disappears in the commutative limit, but nonetheless should be taken into account if one insists
on writing the hCS action as a trace over ${\cal H}_{\mathbb{CP}^{3|4}}(N)$. There are two ways to deal with
this subtlety: (i) stick to the covariant action (\ref{haas}), or (ii) promote
$\Omega_{\alpha \beta \gamma}$ in (\ref{mmactt}) to a dynamical variable that acts like a compensator field.
We will return to this point in section \ref{sec:GRAVFUZZ}.

\subsubsection{Small Phase Space Description}

The twistor matrix model as just given can be viewed as a
large phase space description. It makes the  $su(4|4)$ symmetry manifest,
but at the expense of working with non-square matrices.
There is also a small phase description which sometimes gives a practical alternative.
Introduce three oscillators $\zeta^{i}$ for an affine patch of
$\mathbb{C}^{3}$. On this space, we can restrict to $\mathbb{B}^{3}(N)$,
all states of $\mathbb{C}^{3}$ of level less than or equal to $N$. In
this description, a homogeneous polynomial of degree $d$ in the
remaining $Z$'s is replaced by a polynomial of degree $d$ in the $\zeta$'s.
Starting from the bulk action of equation (\ref{mmactt}), there is a local
direction normal to $\mathbb{CP}^{3}$, which we can identify with $Z^{4}$.
This amounts to the formal replacement of $D^{4}$ and $Z^{4}$ by the identity.
Viewing all fields as polynomials in the $\zeta$'s and $\zeta^{\dag}$'s, with
the maximal degree fixed by their homogeneous counterparts, the
hCS action simplifies to
\bea
S_{\mathrm{hCS}}= \frac{1}{g^{2}} \text{Tr}\left(  \hspace{1pt}\varepsilon_{stu}\hspace
{1pt}D^{s}D^{t}D^{u}\hspace{1pt}\right)  _{\psi^{4}}. \label{sbulk}%
\eea
where the trace is over the three ball $\mathbb{B}^{3|4}$ of states in $\mathbb{C}^3$ with level less or equal to $N$.

\subsection{Defect Action}

We now turn to describe the properties of the defect action (\ref{mmactd}). We first discuss the
free action, with the coupling to the holomorphic Chern-Simons gauge field turned off.
So in (\ref{mmactd}) we replace the gauge covariant derivative ${\cal D}^I$ by the non-commutative
twistor coordinate ${\cal Z}^I$. Also, to keep the discussion a bit more transparent, we will for the
most part restrict our attention to the bosonic sector. The generalization to the supersymmetric formulas is
straightforward.
The free bosonic defect action takes the simple form
\bea
\label{gaussian}
S(\widetilde{Q},Q)\! \is\!  
 \text{Tr}\bigl( \widetilde{Q}\Dbar \smpc Q\bigr) , \\[4mm]
 \label{dbar}
\Dbar Q \, \equiv \!\!\! \! & & \!\!\!\!\!\!\! I_{\alpha\beta}Z^{\alpha} Q Z^{\beta}\text{,}
\eea
where the trace is over the Hilbert space of points on $\mathbb{CP}^{3}$ and over the $u(N_{c})$ color space.
This action defines a gaussian matrix model, where the integration variables $\widetilde{Q}$ and $Q$ are
slightly non-square matrices of size
$k_{N+1}\times k_{N}$, which in addition may carry a color index, transforming respectively
in the $\overline{N_{c}}$ and $N_{c}$ of $U(N_{c})$.

The  gaussian matrix model (\ref{gaussian}) turns out to have quite
remarkable properties, which are studied in some detail in the companion paper \cite{GMM}.
Here our main task is to motivate why (\ref{gaussian}) is the correct action for the defect
modes, that is, for the low energy fluctuations around the homogeneous instanton background.
So let us list the main characteristics of  (\ref{gaussian}), and compare them with our wish list.

\smallskip

$\bullet$ {\it Global symmetries.} Our first requirement is that the defect action should respect the $SO(5)$ symmetry of the
homogeneous instanton background. The action (\ref{gaussian}) clearly does -- but it is instructive to see how it works out.
Let us first look at the global symmetries of  the kinetic operator $\Dbar$ defined in (\ref{dbar}).
The operator $\bar{D}$ is in fact invariant under a full set of $gl(4,\mathbb{C})$ transformations.
Consider a $gl(4,\mathbb{C})$ generator $\mathcal{M}_{\alpha\beta}=Z_{\alpha}^{\dag}Z_{\beta}$,
and define its action on $Q$ via
$\mathcal{M}_{\alpha\beta}^{\, \circ}Q\equiv\mathcal{M}_{\alpha\beta} Q-Q\mathcal{M}_{\beta\alpha}$.
Note that the indices $\alpha$ and $\beta$ switch locations between the two terms. A short
computation shows that $[\Dbar, \mathcal{M}_{\alpha\beta}^{\, \circ}]=0$, and so the kinetic  operator is
invariant under the group generated by all $\mathcal{M}_{\alpha\beta}^{\, \circ}$ operators.
It would not be correct, however, to conclude that the defect action preserves
$gl(4,\mathbb{C})$:  the action (\ref{gaussian}) involves a trace, which is defined with
respect to a choice of inner product. So the most we could hope for is the
symmetry algebra $u(4)$. Additionally, the action contains the infinity bi-twistor $I_{\alpha\beta}$,
which transforms as a vector of $so(6) \simeq su(4)$. This breaks the symmetry group to $so(5)$.
Concretely, hermitian generators that leave the action invariant are
\bea
\label{sofivegen}
\mathcal{M}(v)=v^{\alpha}_{\, \beta} \spc Z_{\alpha}^{\dag}\spc Z^{\beta}, \qquad v^{\alpha\beta}=v^{\beta\alpha},
\eea
where the raising of
the second index is accomplished with the $I^{\beta\gamma}$ bi-twistor. This leaves 10 hermitian charges,
which are the symmetry generators of $so(5)$. In a similar way, we find that the supersymmetric action
(\ref{mmactd}) preserves the symmetries of the supersphere $S^{4|8}$.

\smallskip

$\bullet$ {\it Ultra-locality.} A second desired property is that the defect action should reflect the quantum Hall
intuition, that its excitations are bound to Landau orbits localized along the $S^4$. Alternatively, using
the twistor string theory terminology of  \cite{Witten:2003nn}, we wish to
see that the defect action shares the properties of an effective action for (the ground states of) open strings
 that stretch between the instantonic D-branes and the space-filling D-brane. Both arguments indicate that
the defect action should be ultra-local in space-time. Eqn (\ref{gaussian}) satisfies this beautifully \cite{GMM}.
Consider the position operator $X_{\alpha\beta}$ introduced in eqn (\ref{xdef}). An easy computation
shows that these operators commute with the kinetic operator:
\be
[\Dbar, X_{\alpha\beta}] = 0.
\ee
The same is true for the hermitian conjugate operators $X_{\alpha\beta}^{\dag}$. Hence the kinetic operator
acts within the eigen space (\ref{xeigen}) of $X_{\alpha\beta}$. It is the precise sense in which
the defect action is ultra-local along the $S^4$. So it has the right to be interpreted as the action of
modes that are tied to local defects, that wrap the twistor lines. It is easy to  verify that this result extends to
the supersymmetric case.

\smallskip

$\bullet$ {\it ${\mathbb{CP}^1}$ Propagation.}  This ultra-local property indicates that the defect modes propagate along
the $\mathbb{CP}^{1}$ fiber directions. Continuing to take guidance from twistor string theory,  we would like to see that, when
restricted to a particular fiber, the action reduces to that of a 2-dimensional chiral free field. Via $SO(5)$ symmetry, it
is sufficient to verify this for the twistor line at the origin. So let $\Phi$ be a field localized at the origin, i.e. it is a
linear map on ${\cal H}_{\mathbb{CP}^{3|4}}$ that is made up only from $\pi_a$ and $\pi^\dag_a$ oscillators.
On this subspace, the kinetic operator $\Dbar$ reduces to
\be
\label{cponed}
\Dbar\Phi_{\raisebox{-1.5pt}{{\small $|$}\scriptsize $\mathbb{CP}^1$}} = \varepsilon^{ab}\pi_a  \Phi \pi_b
\ee
Via the commutation relation $[\pi_a,\pi^\dag_b] = \varepsilon_{ab}$, we immediately see the right-hand side indeed acts on $\Phi$
as a standard Dolbeault operator $\bar{\partial} = \pi^a${\large $\frac{\partial\ }{\partial\overline{\pi}^a}$} on $\mathbb{CP}^1$.
Based on this alone, we expect that the propagator of the $\Dbar$ operator looks like the propagator
of a two-dimensional chiral free field. The propagator is studied in some detail the companion paper \cite{GMM},
where this expectation is confirmed. For additional discussion of the Dolbeault operator on fuzzy $\mathbb{CP}^{1}$, see e.g. \cite{Grosse:1994ed, Grosse:1995jt, Dolan:2007uf}.

\smallskip

$\bullet$ {\it ADHM correspondence.} The defect modes $Q$ and $\widetilde{Q}$ have a familiar analogue the ADHM
construction of instantons. From the comparison with ADHM we learn that they naturally transform in the fundamental
and anti-fundamental of $u(N_{c})$. In the full system described by the action (\ref{mmacto})-(\ref{mmactd}),
they provide a source term to the equation of motion of the holomorphic Chern-Simons gauge field.
Looking at (\ref{mmactd}) and the small phase space form (\ref{sbulk}) of the hCS action,
 we can already see that the (top component of) the $D^\alpha$ equation of  motion takes the schematic form of the ADHM equation.
We will make this match more precise in the next section.

\smallskip

$\bullet$ {\it MHV correspondence.} The gauged defect action  is invariant
under gauge rotations%
\bea
Q\rightarrow e^{ih}Q\text{, }\ \widetilde{Q}\!\! & \! \rightarrow\! & \!\! \widetilde{Q}%
e^{-ih}\text{, \ }{\cal D}^{I}\rightarrow e^{ih}{\cal D}^{I}e^{-ih}\text{.}
\label{gaugetransform}%
\eea
We read off that the bulk gauge field couples to the current:
\bea
\label{bil}
J_{\beta}^{A}=T_{ij}^{A}Q^{i}Z_{\beta}\widetilde{Q}^{j}%
\eea
where $T_{ij}^{A}$ is a $U(N_c)$ generator in the fundamental representation. This current is
conserved on-shell: $Z^{\beta}J_{\beta}^{A}=0.$ Hence, after taking the large $N$ limit, the matrix model has all
elements in place to provide a candidate dual description of 4D ${\cal N}=4$ scattering amplitudes. In particular,
following  \cite{Nair:1988bq} and  \cite{Witten:2003nn}, we are led to
identify the correlation functions of these current operators with MHV gluon amplitudes. As shown in \cite{GMM},
the two sides indeed match. We will study this relationship in more detail in section 6.

We have verified that the defect action (\ref{mmactd}) passes several non-trivial checks, which support
its candidacy as the correct action for low energy fluctuations around the instanton background. In the next sections,
we will study the physical properties of the matrix model in more detail. In particular, we will present evidence that, in
a suitable large $N$ limit, its correlation functions approach those of ${\cal N}=4$ gauge theory, coupled to
gravity. Various technical details will be delegated to \cite{GMM}.

\section{ADHM\ Limit} \label{sec:ADHM}

Let us briefly review the realization of ADHM\ in
terms of bound states of branes. The brane construction of ADHM is discussed
in \cite{Witten:1994tz, Douglas:1995bn, Douglas:1996uz} (see \cite{TongRev}
for a concise review).  To frame our discussion, consider type IIB\ string
theory on $\mathbb{R}^{9,1}$ with $n$ D7-branes filling $\mathbb{R}%
^{3,1}\times\mathbb{R}^{4}$, and $k$ probe D3-branes filling $\mathbb{R}%
^{3,1}$. We focus on the $\mathbb{R}^4$ directions, relative to which the D3-branes
look like instantons.

The theory on the probe D3-branes is given by a $U(k)$ $\mathcal{N}\! =\! 2$ supersymmetric field theory.
In $\mathcal{N}\! =\! 1$ language, the motion of the D3-branes is parametrized by
three chiral superfields, $\phi_{1}$, $\phi_{2}$ point parallel  and
$\varphi$ points normal to the seven-brane. All three fields
are in the adjoint of $U(k)$. Additionally, there are bifundamentals
$\widetilde{q}$ and $q$ which define $n\times k$ and $k\times n$ matrices.
These correspond to $3-7$ strings. The superpotential reads:%
\bea
W_{probe}=\text{Tr}_{U(k)}\left(  \varphi\left[  \phi_{1},\phi_{2}\right]
\right)  -\text{Tr}_{U(n)}\left(  \widetilde{q}\varphi q\right)
\label{WPROBE}%
\eea
In this system, the
trace over $U(n)$ is associated with the flavor symmetry of the D7-branes.
There is also a D-term potential. To study the classical vacua, it is enough
to consider solutions to the F-term equations, modulo the
complexified gauge group $GL(k,\mathbb{C})$:
\bea
\left[  \phi_{1},\phi_{2}\right]  =q\widetilde{q}\text{, }\left[  \varphi
,\phi_{2}\right]  =\left[  \varphi,\phi_{1}\right]  =0\text{, }\widetilde
{q}\varphi=\varphi q=0. \label{ADHM}%
\eea
These are the ADHM equations.
The moduli space  consists of several branches: the
Coulomb branch, the Higgs branch, and  mixed branches. The Coulomb
branch, $\widetilde{q}=q=0$, describes the motion of the D3-branes away from
the D7s. On the Higgs branch, $\varphi=0$, and $q$ and $\widetilde{q}$
are non-zero. In this case, the D3-branes have dissolved as instantons
inside the D7-branes. On the mixed branches, some combinations of both are non-zero.

The Higgs branch coincides with the $k$-instanton moduli space of
the $GL(n,
\mathbb{C}
)$ gauge theory on $\mathbb{R}^{4}$. This case is characterized by $\left[
\phi_{1},\phi_{2}\right]  =q\widetilde{q}$, and the space of solutions modulo
$GL(k,\mathbb{C})$ has real dimension $4kn$, which is the dimension of the
instanton moduli space.\footnote{The counting is as follows. We have
$4k^{2}+4kn$ real degrees of freedom from $\phi_{1}\oplus\phi_{2}$ and
$\widetilde{q}\oplus q$. The F-term constraints impose $2k^{2}$ conditions,
and modding out by $GL(k,\mathbb{C})$ removes a further redundancy of $2k^{2}%
$. The resulting moduli space has dimension $4kn$. On $S^{4}$, the instanton
moduli space is $4kn-n^{2}$, but if one allows gauge transformations at
infinity, the dimension is again $4kn$.} More generally, there can be isolated
solutions with all fields switched on. From
the perspective of the D3-branes, such mixed branches describe configurations
where the D3-brane has partially puffed up in the directions normal to the
seven-brane.

Let us now see how this matches up with the behavior of the matrix model. To
do this, we will consider a slightly broader notion of the system, where all
modes are complexified. We can see roughly how the match works.
There are three independent adjoint fields $\phi_1,\phi_2$ and $\varphi$,
which are naturally identified as the covariant derivatives $D^\alpha$.
In a local patch of the $S^{4}$, we can pick a preferred origin and a
corresponding $\mathbb{R}^{4}$. Equipping this with a complex structure
we have a $\mathbb{C}^{2}$, and two corresponding vector fields $D_1$ and $D_2$.
The third component $D_0$ points along the
direction of the twistor line at the origin. In addition, the defect modes
$\widetilde{Q}$ and $Q$ have their analogues with the D3-D7 bi-fundamentals $\tilde{q}$
and $q$.
Note, however, that the size of the matrices do not yet match: $\widetilde{Q}$ and $Q$
are $k_{N+1}\times k_N$ matrices, whereas $\widetilde{q}$ and $q$ are $k\times n$
matrices.

We now provide a more precise match with ADHM. First let us specify the limit in which
we should expect that match to become exact. For this we need to be able to turn off
the non-commutativity. This is done as follows.
Recall the form of the infinity bitwistor:%
\bea
\label{suptwist}
\mathcal{I}_{IJ}=\frac{1}{\ellcc}\left(  \!%
\begin{array}
[c]{ccc}%
\gamma\varepsilon_{\dot{a}\dot{b}} & 0 & 0
\\
0 & \varepsilon^{ab}  & 0\\
0 & 0 & \eta_{ij}\!
\end{array}
\! \right)
\eea
Let us consider the limit $\gamma\rightarrow0$ with $\ellcc$ held fixed.   As explained,
the $\gamma$ parameter describes the ratio of two $S^4$ radii, namely (i) the radius of the $S^4$
relative to which the abelian flux of the Yang monopole is homogeneous, and (ii) the radius of
the $S^4$ relative to which the $SU(n)$ instanton part is homogeneous.
So sending $\gamma \to 0$ amounts to localizing the abelian flux in a small region near the
north pole relative to the non-abelian flux, or equivalently, localizing the non-abelian flux
in a small region near the south pole relative to the abelian flux. From either perspective,
it is clear that in this limit, the effect of the abelian flux on the  non-abelian configuration
becomes negligible. Let us take the perspective where we zoom in on the south pole region of the $S^{4}$.
In this limit, the four sphere starts to look like flat space time, and  moreover, only
the non-abelian part of the configuration is retained.

Varying the matrix action (\ref{mmactd}) with respect to $\widetilde{Q}$,  we obtain the
equation of motion
\bea
\overline{\mathcal{D}}_{\! \mathcal{A}}Q \equiv {\cal I}_{IJ} {\cal D}^I Q {\cal Z}^J = 0 \label{eom}%
\eea
When $\mathcal{I}_{IJ}$ is invertible, this equation generically possesses no solutions. Indeed, via the
rank-nullity theorem, a map $\overline{\mathcal{D}}_{\mathcal{A}}%
:K_{N+1}\times K_{N}\rightarrow K_{N}\times K_{N+1}$ will in general have a
trivial kernel. In the specific limit $\gamma\rightarrow0$, however, $\mathcal{I}_{IJ}$ is no
longer invertible and the matrix system will have a moduli space of solutions.  Looking at eqns (\ref{suptwist}) and (\ref{eom})
we see that derivatives in the $\omega$ direction cost much less energy than
derivatives in the other directions of the supertwistor space. The effective
geometry experienced by the modes is $\mathbb{C}^{2|4}\times\mathbb{CP}^{1|0}$
where the $\mathbb{CP}^{1|0}$ factor is composed from the $\omega$ directions.
(In the presence of a background gauge field, the overall direction of this
constraint will change, but this can be absorbed into the definition of the
oscillators.) Without loss of generality, the constraint is then:
\bea
\pi_{a}Q=\psi^{i}Q=0
\eea
These conditions project $Q$ down to a $(n+1)\times K_{N}$ supermatrix, which is purely
bosonic in the $(n+1)$-component part. In other words, the left index on $Q$ are states of a bosonic
$\mathbb{CP}^{1|0}$ at level $n$. The kinetic term for these modes then fixes
the form of $\widetilde{Q}$, so that it is given by a $K_{N+1}\times
n$ supermatrix.

In this limit, it is also appropriate to take a specific basis for the bulk
gauge fields. There are the two derivatives in directions transverse to the
$\mathbb{CP}^{1|0}$, given by $D^{1}$ and $D^{2}$. In addition, there is the
derivative along the $\mathbb{CP}^{1|0}$, given by $D^{0}$. To analyze the
remaining equations of motion, it is helpful to pass to a description in terms
of an affine patch. This leaves us with three matrices $D^{0}$, $D^{1}$,
$D^{2}$ which admit a $\psi$-expansion in terms of $k_{N}\times k_{N}$ bosonic
matrices, and $Q$ and $\widetilde{Q}$ which respectively admit $\psi$- and
$\psi^{\dag}$ expansions in terms of bosonic matrices of respective sizes
$(n+1)\times k_{N}$ and $k_{N}\times(n+1)$. The resulting action is:
\bea
S=\text{Tr}_{{}_{K_N}}\!\! \left(  D^{0}\left[  D^{1},D^{2}\right]
\right)  _{\psi^{4}}+\text{Tr}_{n}\bigl(  QD^{0}\widetilde
{Q}\bigr)
\eea

The ADHM equations now naturally appear by varying
with respect to the $\psi^{4}$ component of $D^{0}=d^{0}+...+\psi^{4}b^{0}$.
First, observe that $\psi^{4} b^0$ is sandwiched between the
supermatrix parts of $Q$ and $\widetilde{Q}$. Since $\psi^{4}$ corresponds to
an annihilator operator when acting to the right on $\widetilde{Q}$, the only
non-zero coupling involves the $\psi^{\dag4}$ component of $\widetilde{Q}$,
which we denote by $\widetilde{q}$. On the other side, the only term that
survives is $q$, the purely bosonic component of $Q$. The terms of
the bosonic action involving $b^{0}$ are then:
\bea
S=\text{Tr}_{\mathbb{CP}^{3}}\left(  b^{0}\left[  d^{1},d^{2}\right]  \right)
+\text{Tr}_{\mathbb{CP}^{1|0}}\left(  qb^{0}\widetilde{q}\right)
\eea
where $d^{i}$ is the bosonic component of $D^{i}$. Setting to zero all other
components of the supermatrices, we see that the resulting equations of motion
are \textit{identical} to those of line (\ref{ADHM}). Working modulo
$GL(k_{N},\mathbb{C})$ transformations, we obtain a moduli space of instantons
for $su(n+1)$ gauge theory at instanton number $k_{N}$. Note that this is the
maximal instanton number for which it is possible to define an $so(5)$
homogeneous instanton configuration:%
\bea
k_{N}=\frac{(N+1)(N+2)(N+3)}{6}=\frac{(n+1)\left(  \left(  n+1\right)
^{2}-1\right)  }{6}.
\eea
This is again in accord with the fact that we are considering instanton
configurations which are deformations of the non-abelian part of the Yang monopole.

\section{Continuum Limit\label{sec:LRAD}}

Our discussion so far has focussed on the finite $N$ regime of the matrix
model.  The finite matrices furnish a basis of functions on  $S^{4}$ but only
up to some cutoff angular momentum. To arrive at a continuum theory,
it is necessary to take a large $N$ limit.

In addition to $N$, the matrix model comes with a
length scale, given by the $S^4$ radius $\ellcc$, and a parameter $\gamma$ which controls the breaking
of the $so(5)$ symmetry. Depending on the scaling of these parameters, we
arrive at different limits for 4D physics. The most basic large $N$ limit is taken
while keeping $\gamma=1$, which preserves the $so(5)$ symmetry.
Alternatively, we can take a combined large $N$ and flat space limit, which
corresponds to performing a Wigner-In\"{o}n\"{u} contraction of the original $so(5)$
algebra. This involves rescaling the momentum and conformal boost generators,
or equivalently, rescaling the $\pi^{\dag}$ oscillator relative to the
$\omega^{\dag}$ oscillator.

We can consider two types of flat space continuum limits. In the first scenario, we
first send $N$ to infinity, and then take the flat space limit $\ellcc \to \infty$. As we will
argue, this continuum limit produces pure ${\cal N}\! =\! 4$ SYM theory.
The correspondence arises in a similar way as in twistor string theory,
and the dictionary looks almost identical.
But there are some important contrasts,
since our treatment of the defect system is quite different from that in \cite{Witten:2003nn}.
Most significantly, we will present clear evidence that in our set-up,
the continuum ${\cal N}=4$ gauge theory arises {\it without} any coupling to conformal gravity.

A second possibility is to take a double scaling limit, in which both $N$ and $\ellcc$ are
sent to infinity, but such that the ratio $\ell_{pl}^2 = \ellcc^2/N$ is held fixed:
\bea
N\to \infty , \ellcc \to \infty, \quad {\rm with} \ \ \ell_{pl}^{2}=\frac{\ellcc^{2}}{N} \ \ {\rm fixed}\, .%
\eea
We will argue that in this limit, the system does contain gravity. Moreover, we will
present evidence that the emerging gravitational theory is described by the usual
Einstein action, with a finite Newton constant $G_N$ of order $\ell_{pl}^2$. The strongest
evidence in support of this inference is that the matrix model is able to reproduce the
MHV graviton amplitudes. The calculation that leads to this result is described in the
companion paper \cite{GMM}. In the next section we will try to give a more conceptual explanation
of how and why the gravitational degrees of freedom can arise from a matrix model.

\smallskip

\subsection{Emergent Gauge Theory}

Let us first begin with a discussion of the purely gauge theory sector. In the
matrix model, this is arranged by taking $\ell_{pl}\rightarrow0$. Here we will
study this limit directly at the level of the action.

The basic dictionary is relatively simple and standard. Starting from the matrix model
action (\ref{mmacto})-(\ref{mmactd}), the first step is to rescale all twistor coordinates
by a factor of $1/\sqrt{N}$, so that the commutation relations take the form $[Z^\alpha, {Z}^\dag_\beta ] = \delta^{\alpha}_\beta/N$,
etc. Hence, upon taking the large $N$ limit, the twistor coordinates become ordinary commutative variables.
Accordingly, the trace of the Hilbert space ${\cal H}_{\mathbb{CP}^{3|4}}$ of points, reduces to an integral over the commutative
supertwistor space $\mathbb{CP}^{3|4}$.  The covariant derivatives $D^\alpha$, instead must be defined with an extra
prefactor of $N$, so that $Z^\alpha$ reduces to a anti-holomorphic derivative $\mbox{\large $\frac{\partial \ }{\partial \overline{Z}_\alpha}$}$.
We will thus write
\bea
D^\alpha = N Z^\alpha +  {\cal A}^\alpha
\eea
 where ${\cal A}^\alpha$ denotes the  gauge field.
 Applying this dictionary to the matrix version of the holomorphic Chern-Simons action gives the continuum hCS action, in the form explained
at the end of subsection \ref{sec:MatCS}. In particular, the stationary points of the hCS action are holomorphic flat connections which satisfy (\ref{holo}.

The appearance of a continuum field theory in the large $N$ limit, while standard and reasonable,
in fact needs a bit more justification. For any finite $N$, regardless how large, the space of functions  on a non-commutative
space includes arbitrarily non-local maps. Indeed, the matrix model involves integration over matrices
$D^\alpha$, $\tilde{Q}$ and $Q$, which include maps from one point to any other point on the fuzzy $\mathbb{CP}^{3|4}$.
In our context,  these non-local maps are exponentially suppressed, due to the specific form of the matrix action.
For example, as seen from (\ref{holo}), the saddle points of the matrix hCS action are maps that are locally holomorphic
in the $Z^\alpha$'s.\footnote{Functions or sections of degree $p$ line bundles on $\mathbb{CP}^{3|4}$ can typically not
be globally holomorphic, since it should be expandable  as a degree zero (or degree $p$) polynomial in terms of
$Z^\alpha$ and $Z^\dag_\alpha$. A function $f(Z)$  locally holomorphic is if its commutator with the holomorphic
coordinates $Z^\alpha$ is an operator that annihilates a subspace ${\cal H}(S)$ of the Hilbert space ${\cal H}_{\mathbb{CP}^3}(N)$.
This subspace ${\cal H}(S)$ represents the local region within which we can consider $f(Z)$ as a holomorphic function of
the $Z$ coordinates.  For a more precise definition of locally holomorphic functions, see  \cite{GMM}}  Since the
$Z^\alpha$'s commute among each other, holomorphic functions multiply locally.
In other words, all non-locality is associated with the $Z^\dag_\alpha$ dependence of the fields. As we will see more
explicitly in the next section, the size of this non-locality is set by $\ell_{pl}$. So in the $\ell_{pl} \to 0$ limit,
the theory becomes local.

The correspondence with local 4D physics proceeds by projecting
the $\mathbb{CP}^{3|4}$ gauge field ${\cal A}^\alpha$ down onto $S^4$. This is done via
the Penrose correspondence,
which repackages the components of ${\cal A}^\alpha$ into
those of ${\cal N}=4$ SYM theory. Adjusted to our context, the Penrose transform works as follows.\footnote{In commutative twistor
theory, the basic correspondence is between space-time fields of helicity $h$,
and certain cohomology elements on $\mathbb{PT}^{\prime}$, that is,
$\mathbb{CP}^{3}$ with the $\mathbb{CP}^{1}$ at infinity removed:%
\bea
\left\{  \text{helicity }h\text{ space-time fields}\right\}  =H^{1}\left(
\mathbb{PT}^{\prime},\mathcal{O}(2h-2)\right)
\eea
where this is defined with respect to a sheaf cohomology. Consider a
meromorphic $(0,1)$ form defined over commutative twistor space, which we
denote by $\Psi(\omega,\pi)$. Imposing the constraint $\omega^{\dot{a}%
}=ix^{\dot{a}a}\pi_{a}$, we obtain a space-time dependent $(0,1)$ form
$\Psi(ix\pi,\pi)$. Given this, we obtain a helicity $+n$ or $-n$ free field on
ordinary space-time via the Penrose transform:%
\begin{align}
\phi_{a_{1}...a_{n}}(x)  &  =%
{\displaystyle\oint_{\gamma}}
\epsilon^{bc}\pi_{b}d\pi_{c}\times\pi_{a_{1}}...\pi_{a_{n}}\Psi\\
\phi_{\dot{a}_{1}...\dot{a}_{n}}(x)  &  =%
{\displaystyle\oint_{\gamma}}
\epsilon^{bc}\pi_{b}d\pi_{c}\times\frac{\partial}{\partial\omega^{\dot{a}_{1}%
}}...\frac{\partial}{\partial\omega^{\dot{a}_{n}}}\Psi
\end{align}
where the integration is over a contour $\gamma$ defined over the
$\mathbb{CP}^{1}$ in the $\pi$ coordinates.}

Suppose ${\cal A}$ is an on-shell mode. So on a suitable subset of the Hilbert space, it commutes with the $Z^\alpha$ coordinates.
In particular, it preserves the twistor line equations (\ref{bosonicline}) that select the $N+1$ dimensional Hilbert space ${\cal H}_x$ associated with a given
space-time location $x$. (Here and below $x$ is short-hand for a point $(x,\theta)$ on the supersphere $S^{4|8}$.)  We can therefore
decompose ${\cal A}$ as a sum of operators
\bea
\label{am}
{\! \cal A}^\alpha  =  \int \! d^{4|8} x \; {\!{ \cal A}^\alpha(x)}
\eea
where ${\!{ \cal A}^\alpha(x)}$ denotes a linear map from ${\cal H}_x$ to ${\cal H}_x$.
As motivated in more detail in \cite{GMM},
an on-shell mode corresponds to taking ${\!{ \cal A}^\alpha(x)}$ of the special form
\bea
\label{penrose}
{\!{ \cal A}^{\alpha}(x)} = \oint \la \lambda d\lambda\ra \, A^{\alpha}(x, \lambda) \spc | x, \lambda ) (x, \lambda|
\eea
where, in order for ${\cal A}$ to represent a holomorphic function of the twistor coordinates $(\omega^{\dot a},\pi_a)$, $A$
must be a function of the form $A(ix^{\dot a a}\lambda_a, \lambda_a)$.  Equation (\ref{penrose}) is the quantum version
of the Penrose transform. The classical version is obtained by sandwiching both sides between the two position eigenstates $(x,0|$ and $|x,0)$,
and using that $(x,0|x,\lambda) = (x, \lambda |x, 0 ) = 1$.

Applying this reduction procedure to the pure hCS theory only gives
the self-dual gauge theory on $S^4$.  The role of the defect modes is to supplement the rest of the ${\cal N}=4$ SYM theory.
At the  level of perturbation theory, the match proceeds via the CSW rules for constructing scattering amplitudes \cite{CSW}.
To see how this emerges from the large $N$ matrix model, let us look more closely at the defect contribution.

As we have seen,  the defect action is ultra-local in space-time. The fields $Q$ and $\tilde{Q}$
do not propagate along the space-time directions, and hence do not correspond ordinary
space-time fields.  The defect modes appear quadratically in (\ref{mmactd}). So it is natural to
integrate them out, and  collect their contribution to the effective action for the space-time gauge field
${\cal A}$ in the form of a functional (or rather, for us,  just a regular) determinant.

Let us write the defect action (\ref{mmactd}) in the short hand notation%
\bea
\label{ndefect}
S_{\text{defect}}= \text{Tr}\bigl(\spc\widetilde{Q}\Dbar_{\! \mathcal{A}}Q\spc \bigr)\ \ & ;  & \ \
\Dbar_{\! \cal A} =  \Dbar +  {\cal A}
\eea
with $\Dbar$ the free kinetic operator defined in eqn (\ref{dbar}).
The trace is over the Hilbert space ${\cal H}_{\mathbb{CP}^{3|4}}(N)$ of point on twistor space. The (lowest superfield components of the)
defect modes $\widetilde{Q}$ and $Q$ are both $k_{N+1}\times k_{N}$ matrices. In the following, the space of all such matrices will be denoted by $M_k$.
Now consider the operator $\Dbar^{-1}\!\! \circ\!  \Dbar_{\! \cal A} = \mathbf{1} +  \Dbar^{-1} {\! \cal A}$.
This defines a map from $M_k$ to $M_k$. So we can consider its determinant.
Integrating out the defect modes leads to an ${\cal A}$-dependent effective action
\bea
\label{seff}
S_{\text{eff}}({\cal A})\is 
\text{Tr}_{\strut M_k}\! \log\bigl(\mathbf{1} +  \Dbar^{-1} {\!\! \cal A} \bigr),
\eea
where we used the standard identity $\log \det = {\rm Tr} \log$.
We will now rewrite this in the form of a generating function of MHV
gluon amplitudes.

Let us specialize to the case that ${\cal A}$ describes a sum of on-shell modes, of the form (\ref{am})-(\ref{penrose}).
The defect mode kinetic operator $\Dbar$ and the  on-shell field $\mathcal{A}$
then both commute with the position operators $X_{\alpha\beta}$
\bea
[\Dbar, X_{\alpha\beta}] = 0 & ; & [\mathcal{A}, X_{\alpha\beta}] = 0\, .
\eea
The operator inside the trace in (\ref{seff}) also commutes with $X_{\alpha\beta}$, and thus defines an ultra-local operator
on space-time. Given this result, we can naturally decompose the trace over $M_k$ into two factors: a trace over the space-time
directions, and a trace ${\rm Tr}_x$ over the space of maps that act on the $N+1$ dimensional Hilbert space ${\cal H}_x$
associated with the space-time point $x$. In the large $N$ limit, this decomposition amounts to
making the replacement
\bea
\label{replace}
\text{Tr}_{M_K}\rightarrow \int d^{4|8}x\text{ Tr}_{\raisebox{-2pt}{\scriptsize$x$}}. \label{measuresub}%
\eea
Here we have made use of the fact that the overlap between states located at different space-time points
takes the form of a sharply peaked gaussian (\ref{overlap}), which in the large $N$ limit turns into a delta-function.
In the continuum theory, (\ref{replace}) simply amounts to writing an integral over $\mathbb{CP}^{3|4}$ as
an integral over the $S^{4|8}$ base, times an integral over the $\mathbb{CP}^1$ fiber. The $\Dbar$ operator
in (\ref{leff}) takes the form (\ref{cponed}), which in the large $N$ limit becomes the standard Dolbeault operator $\overline\partial$ on $\mathbb{CP}^1$.
In the companion paper \cite{GMM}, we work out the precise form of the Green's function $\Dbar^{-1}$ and show that at large $N$
it coincides with the continuum version. We can view $\Dbar$ as the kinetic operator of a free chiral field theory in two dimensions.
To reflect this, we shall make the replacement $\overline{D} = \overline{\partial}$ when acting on a given twistor line.

We thus conclude that  (\ref{seff}) in fact describes a local space-time action
\bea
\label{nseff}
S_{\text{eff}}({\cal A}) = \int \!\! d^{4|8} x \spc {\cal L}_{\rm eff}({\cal A}(x))
\eea
The effective Lagrangian takes the form
\bea
\label{leff}
{\cal L}_{\rm eff}({\cal A}(x)) \is \text{Tr}_x 
\log\Bigl(\mathbf{1}\! \smpc + \overline{\partial}^{-1} {\!\! \cal A}(x)\Bigr) \\[3.5mm]
\is\; \log\text{det}_x 
\bigl(\mathbf{1}\! \smpc +\overline{\partial}^{-1} {\!\! \cal A}(x)\bigr)
\eea
where ${\rm Tr}_x$ and $\det_x$ respectively denote the trace and determinant over the space of linear maps on ${\cal H}_x$, or in more geometric terms,
the space of functions on the commutative limit of a fuzzy $\mathbb{CP}^{1|0}$ twistor line located at $x$. Equation (\ref{leff}) is then the generating function
1PI correlation functions of bi-linear currents of the chiral free fields, of the form (\ref{bil}). In \cite{GMM} we show that the large $N$ limit of
these current correlation functions reproduces the Parke-Taylor formula.

At this point we have made contact with the works \cite{Boels:2006ir, Boels:2007qn}, where it is shown that
the continuum version of the effective action (\ref{nseff})-(\ref{leff}) is the generating function of gluon
MHV amplitudes. Importantly, as pointed out in these references, the effective action (\ref{leff})
evades the troubling appearance of conformal gravity amplitudes, that plague the twistor
string proposal of \cite{Witten:2003nn, Berkovits:2004jj}. The main difference between the two
proposals is that in the twistor string theory, the effective action that gave the MHV amplitudes
arose from integrating out the open string modes attached to defect D1-instantons, which then
were integrated over their moduli space of positions. This latter procedure gives an effective
lagrangian of the form $\det\bigl(\mathbf{1} + \overline{\partial}^{-1}{\cal A}\bigr)$, rather than
$\log\det\bigl(\mathbf{1} + \overline{\partial}^{-1}{\cal A}\bigr)$.
The $\det$ expression of twistor string theory has several drawbacks. First, it is not by
itself gauge invariant, and to be well-defined, requires additional couplings to a closed string sector.
By contrast, the $\log\det$ term is gauge invariant. In the continuum theory, this is due to a subtle interplay
with $\mathcal{N}=4$ superspace.\footnote{Indeed, without the superspace integrations, the MHV action of  \cite{Boels:2006ir, Boels:2007qn}
would not have been gauge invariant, and moreover, would have contained additional divergences due to ultra-local
interactions and accompanying factors of $\delta(0)$.} For us, the gauge invariance of (\ref{leff}) is manifest, since it was derived by integrating
out a regulated and manifestly gauge invariant action (\ref{ndefect}).

Let us note that when there is reduced supersymmetry, the cancellation of chiral anomalies is a bit
more subtle.  In this case, the 2D chiral determinant in (\ref{leff}) would produce an anomaly outflow, that
somehow needs to get cancelled by some anomaly sink somewhere else.  The appearance of the
anomaly is related to subtleties with the analogue of equation (\ref{measuresub}). With less than
${\cal N}\!=\! 4$ supersymmetry,  the $\mathbb{CP}^3$ Hilbert space does not neatly factorize as
assumed in (\ref{measuresub}). The obstruction is the Berry phase associated with
passing from one coherent state $\left\vert x\right)  $ to another $\left\vert x^{\prime}\right)  $.
The associated Berry connection is the Yang monopole background. The Yang monopole
background transforms under gauge transformations, and in this way, is capable of
transporting the anomaly from one twistor line to another.  So we see that the matrix model
regulates the $\log\det$ action proposed in \cite{Boels:2006ir, Boels:2007qn}, and should
make sense with reduced supersymmetry.

Returning to the ${\cal N}\! = 4$ symmetric case, one could perhaps wonder how it could be possible that in the $\ell_{pl} \to 0$ limit,
one can recover a superconformally invariant continuum theory. At any finite $\ell_{pl}$, only the $SO(5)$ symmetry is unitarily realized
in the matrix model, so where could the other generators come from? The mechanism is expected to be the same as for any non-conformal theory
with a (non-trivial) IR fixed point: conformal symmetry is broken as long as the UV scale is present, and arises as a linearly and
unitarily realized symmetry only in the strict IR limit.

Finally let us comment on non-MHV amplitudes. The basic story here is the same
as in the continuum theory. The total matrix action combines
the holomorphic Chern-Simons action with the effective action (\ref{nseff})-(\ref{leff}).
As in \cite{CSW, Adamo:2011pv}, general types of amplitudes are obtained by
gluing  together MHV amplitudes via propagators of the hCS theory. Here we do not have much
to add to the continuum discussion, except for the folllowing comment: instead of integrating
out the defect modes, we can alternatively integrate out the bulk modes. Specifically,
one can choose an axial gauge\footnote{Another natural and potentially useful gauge condition on ${\cal A}$ is
$Z^\alpha \mathcal{A}_\alpha = 0$. Note that this condition involves the infinity twistor.}, for which the interaction
terms of the hCS theory drop out. So we can just perform the gaussian integral over the gauge field.
The effective action for the defect field $Q$ and $\tilde{Q}$ will then contain the kinetic term, and a non-trivial quartic interaction term, which can be included iteratively. It is tempting to conjecture that the BCFW recursion
relations \cite{Britto:2005fq} can be interpreted as the loop equation of this interacting matrix model.
We leave the exploration of non-MHV and loop amplitudes for future study.

\section{Emergent Gravity \label{sec:GRAVFUZZ}}

In the previous sections we have provided evidence that the continuum limit of
the matrix model specifies a $U(N_{c})$ gauge theory on an $S^{4}$. We have
also seen signs that hint at the presence of gravity.
This is reflected in the  appearance of a minimal length scale $\ell_{pl}$,
and the fact that gauge transformations on the non-commutative
space look like diffeomorphisms.

In this section we assemble these hints into concrete evidence that the matrix model
contains a gravitational subsector. In the
companion paper \cite{GMM} we show that MHV\ graviton scattering amplitudes
can be represented as correlation function of currents of the gaussian matrix model, that describes
the defect modes in a fixed ${\cal A}$ background. In
this section, we will try to embed this result inside a more global geometric perspective.
The appearance of a gravitational subsector is in some sense anticipated by earlier work on
twistor string theory. However, as opposed to there, the matrix model comes
with an intrinsic length scale, and therefore does not seem to describe
conformal gravity. We comment on this point further in subsection
\ref{ssec:CONF}. In this section, we will put $N_c=1$.

\subsection{Currents and Compensators}

In this subsection we show that the matrix model contains gravitational current operators.
To some extent, the presence of this sector is necessary in order for the
non-commutative Chern-Simons action to even be well-defined.  As explained earlier,
starting from the hCS action (\ref{haas}), written as an integral over the
four-ball $\mathbb{B}^{4|4}$, one could in principle perform a partial integration
and obtain an action of the form (\ref{mmactt}). However, the naive action (\ref{mmactt})
is not invariant under the gauge transformations $D^{\alpha}\rightarrow e^{ih}D^{\alpha}e^{-ih}$,
since the $h$'s do not commute with $\Omega_{\alpha \beta \gamma}$. To rectify this
problem, we can introduce a compensator field
$\Omega^{\alpha} = \frac{1}{6} \varepsilon^{\alpha \beta \gamma \delta} \Omega_{\beta \gamma \delta}$
which transforms in the adjoint under non-commutative $U(1)$ gauge transformations. The
correct action is then:%
\begin{equation}
S_{\text{fhCS}}= \frac{1}{g^2} \varepsilon_{\alpha\beta\gamma\delta}\text{Tr}\left(
\Omega^{\alpha} D^{\beta}D^{\gamma}D^{\delta}\right)  .
\end{equation}
One can view the $\Omega^{\alpha}$ as a residual boundary  or `singlet'  component of the
$U(N_c)$ gauge field  on the eight dimensional ball $\mathbb{B}^{4|4}$, which is left
over after performing the partial integration.

This gives a clear indication that the adjoint $U(1)$ factor of the fuzzy
system is on a different footing from the other gauge symmetries. In
commutative terms, maintaining gauge invariance via the compensator can be
interpreted as allowing the holomorphic form $\Omega=\varepsilon_{\alpha
\beta\gamma\delta}Z^{\alpha}dZ^{\beta}dZ^{\gamma}dZ^{\delta}$ to be deformed,
akin to a Kodaira-Spencer theory of gravity \cite{Bershadsky:1993cx}.
In so doing, the geometric data of the twistor space become dynamical.
This already suggests the appearance of a gravitational subsector.

It is reasonable to assume that the compensator should also enter in the defect action.
Indeed, there is a natural place for it. Viewing the defect action as descending from a holomorphic
Chern-Simons action in the presence of a flux defect (as is common in the study of intersecting brane
configurations), the compensator appears in the action as:%
\bea
S_{\text{defect}}\is \text{Tr}\left(  I_{\alpha\beta}\widetilde{Q}D^{\alpha
}Q \Omega^{\beta} \right)\, .
\eea
The $Q$ and $\widetilde{Q}$ modes have become bifundamentals of $u(N_{c})\times u(1)$. The system is
invariant under the enlarged group of gauge transformations:%
\begin{equation}
Q\rightarrow e^{ih}Qe^{-ig}\text{, \ \ }\widetilde{Q}\rightarrow
e^{ig}Qe^{-ih}\text{, \ \ }D^{\alpha}\rightarrow e^{ih}D^{\alpha}%
e^{-ih}\text{, \ \ } \Omega^{\alpha} \rightarrow e^{ig}\Omega^{\alpha}e^{-ig}.
\end{equation}
where $h$ is a Lie algebra element of $u(N_c)$ and $g$ is a phase, and both are local functions of $Z$ and $Z^\dag$.
In addition to the usual color gauge rotations, we see that the
gauge transformations can also shift the locations of fuzzy points. In other
words, the full symmetry is $u(N_{c})\times u(1)\times gl(k_{N}%
)\times\widetilde{gl}(k_{N})$ where the two $gl(k)$ factors act by left and right
multiplication on the defect fields.
In keeping with the interpretation of the $\Omega^{\alpha}$ mode as a
compensator field,  it is natural to focus on the diagonal
$gl(k_{N})_{diag}\subset gl(k_{N})\times\widetilde{gl}(k_{N})$. Indeed, if we
view the matrix $Q$ as a field located at some holomorphic point $x$, then the simultaneous
left and right action by the diagonal $gl(k_{N})_{diag}$
corresponds to a diffeomorphism on the 4D
space-time which is capable of shifting $x$ to another location.

Following the Noether procedure, we obtain the $gl(k_{N})\times\widetilde
{gl}(k_{N})$ `currents' (here we switched notation, and explicitly write the $gl(k_N)$ matrix incdices)
\bea
\label{gravcurrents}
J_{\beta}(T) =\mathcal{T}_{a\overline{b}}Q^{a\overline{c}}Z_{\beta}\widetilde{Q}^{c \overline
{b}}\text{, \ \ \ }\widetilde{J}_{\beta}(\widetilde{T}) =\mathcal{\widetilde{T}}_{a\overline{b}}%
\widetilde{Q}^{a\overline{c}}Z_{\beta}Q^{c \overline{b}}.
\eea
where $\mathcal{T}_{a\overline{b}}$ is a generator of $gl(k_{N})$. The diagonal currents are the linear combinations%
\begin{equation}
\mathfrak{J}_{\beta}(T)=J_{\beta}(T)-\widetilde{J}_{\beta}(T).
\end{equation}
In principle, one can consider charges associated with arbitrary $gl(k_N)$
transformations, involving an arbitrary number of $Z^\dag$ oscillators.
However, we can anticipate that the equations of motion of the corresponding bulk
gauge field disfavors such anti-holomorphic dependence. So it is reasonable
to restrict our attention to currents of the form (\ref{gravcurrents}) which contain
only one single $Z^\dag$ oscillator. We will write out these currents in the
next subsection.

Associated with the diagonal $gl(k_N)$ is a gauge field
$\mathfrak{A}^{\beta}$, which is a function of $Z$ and $Z^\dag$.
It couples to the defect system via the vertex
operator:%
\bea
\mathcal{T}(\mathfrak{A})  \is \text{Tr}\left(  \mathfrak{A}^{\beta
}J_{\beta}\right)  -\text{Tr}(\mathfrak{A}^{\beta}\widetilde{J}_{\beta
})\label{defectcoup}%
\eea

Based on the index structure of the diagonal $gl(k_{N})$ charges (\ref{gravcurrents}), we see that
the symmetries define $(0,1)$-forms valued in the holomorphic cotangent and
tangent bundle to twistor space. This is the same mode content expected for a conformal
graviton \cite{Berkovits:2004jj}, but with a few important differences. First,
we see that they couple to a defect action which involves the infinity
bitwistor. This means that although these currents are obtained from conformal
gravitons, they are subject to the constraint that they leave the defect
action, and thus the infinity twistor invariant. This imposes
conditions which truncate the physical mode content.
The combined presence of a UV cut-off and of the infinity twistor both break conformal invariance,
which suggests that the emerging gravitational sector will not be conformal.

We can extend our discussion to the supersymmetric case with defect action:%
\begin{equation}
S_{\text{defect}}=\text{Tr}\left(  \mathcal{I}_{IJ}\widetilde{Q}%
\mathcal{D}^{I}Q \Omega^{J}\right)\, .
\end{equation}
The bosonic currents (i.e. with $I$ and $J$
bosonic) provide us with the gauged translations and $u(N_{c})$ rotations. The currents with $I$ and $J$ both fermionic give rise to
$su(4)$ gauge fields.
Further,  there is a fermionic current as well which couples to the gravitinos.
All this is in accord with the gauge symmetries of an $\mathcal{N}=4$ supergravity
theory.\footnote{The bulk gauge field $\mathcal{A}^{I}$ also can be viewed as a
vector on superspace. We are making the simplifying assumption that these
additional components of the would be gauge field in the fermionic directions
have been set to zero. It would be interesting to study this more general
situation, though it is unclear to us that it is a necessary element of the construction.}

\subsection{MHV\ Graviton Scattering}

To confirm the proposed geometric interpretation of the $gl(k_{N})$ symmetry, in
the companion paper \cite{GMM} we show that correlation functions of the currents (\ref{gravcurrents})
reproduce MHV graviton scattering amplitudes. Here we just briefly comment on how this
structure emerges from the matrix model, and on how it relates to existing results obtained
from ordinary twistor space.

MHV amplitudes in gravity involve a single ingoing negative helicity graviton
and an arbitrary number of ingoing positive helicity gravitons. Collectively,
the plus helicity gravitons  represent a selfdual
background geometry, on which the minus helicity graviton propagates.
The MHV amplitudes arise by expanding out
the background field in terms of linearized perturbations around flat space-time.
The twistor realization of this calculation builds on Penrose's non-linear
graviton \cite{Penrose:1976jq}. In this construction,
(anti-)selfdual space time backgrounds are represented as complex structure deformations of twistor space.
Infinitesimal deformations correspond to vector fields, that via their Lie derivative,
bend the location of the holomorphic twistor lines. As shown in \cite{Mason:2008jy},
this geometric description allows a computation of MHV graviton amplitudes, that
reproduces the well-known BGK formula \cite{Berends:1988zp}.

The basic setup for scattering theory in the matrix model is to zoom in on a
small and locally flat patch near the south pole of the $S^{4}$. This allows us to define a
flat space limit via the Wigner-In\"on\"u contraction of the $so(5)$ algebra:%
\begin{equation}
\label{calp}
\mathcal{P}\rightarrow P+\ellcc^{-2}K
\end{equation}
so that locally, translations are given by just the $P$. As shown in \cite{GMM},
the asymptotic data of the scattering theory  are specified by  vertex operators $T$ such
that the adjoint $U(1)$ gauge field takes the special form $\mathfrak{A}^{\beta}=Z^{\beta}T$,
where $T(Z,Z^\dag)$ is some non-commutative function. The $gl(k_{N})$ generators
of interest for MHV graviton scattering are of the form which locally have a single $Z^{\dag}$ oscillator:
\begin{equation}
T_{v}(p)=(v\cdot\mathcal{P})\Psi(p)
\end{equation}
where $v^{\dot{a}a}$ is a polarization tensor for the graviton,
$v\cdot\mathcal{P}=v^{\dot{a}a}P_{\dot{a}a}+v_{\dot{a}a}K^{\dot{a}a}$ is a
complexified $so(5)$ generator, and $\Psi(p)$ is a \textquotedblleft momentum
eigenstate\textquotedblright\ matrix satisfying $[P_{\dot{a}a},\Psi
(p)]=p_{\dot{a}a}\Psi(p)$ for complexified momentum $p_{a\dot{a}}=\lambda
_{a}\widetilde{\lambda}_{\dot{a}}$.
The corresponding currents for the plus helicity gravitons build up a self-dual background, and are given by:
\bea
\mathcal{T}_{+} =\text{Tr}\Bigl(\bigl[T,Q\bigr]\spc \overline{D}\widetilde{Q}\Bigr)
\eea
the minus helicity gravitons are specified by the related set of currents:
\bea
\mathcal{T}_{-} =\text{Tr}\Bigl(T Q \spc \overline{D}\widetilde{Q}\Bigr)
\eea
which are acted upon by the $\mathcal{T}_{+}$ generators.
In an $\mathcal{N}= 4$ supersymmetric theory these modes sit in different supermultiplets, and so can naturally be associated
with different types of currents. Quite remarkably, the
correlation functions of these currents reproduce the MHV graviton scattering amplitude of Einstein gravity in flat space-time.
This correspondence is worked out in detail in \cite{GMM}, and forms the main piece of evidence
that  our large $N$ twistor matrix model gives rise to a space-time theory that contains
gravity.

\subsection{Geometric Action}

In light of the detailed evidence that the defect sector of the matrix model contains
the MHV sector of gravity, it is appropriate to ask whether the result can be extended to
include non-MHV dynamics. Following the gauge theory lead, the logical place to look is
in the hCS gauge sector and its coupling with the gravitational currents of the
defect system. Indeed, we have already seen hints of a gravitational mode
in the hCS system, in the form of an obstruction to decoupling the adjoint $u(1)$
gauge transformations.\footnote{Though the physical context is different,
for somewhat related discussion on non-commutative gauge theory and emergent gravity,
see \cite{Rivelles:2002ez, Yang:2004vd, Yang:2006mn, Steinacker:2007dq, Steinacker:2008ri}
and \cite{Steinacker:2010rh} for a review.}

To discuss the action of the adjoint $u(1)$ in the large $N$ limit, it is
helpful to reformulate the matrix multiplication of the matrix fields in terms
of a Moyal product on the homogeneous coordinates:%
\begin{equation}
Z^{\alpha}\ast\overline{Z^{\beta}}=Z^{\alpha}\overline{Z^{\beta}}%
+\varrho^{\alpha\overline{\beta}}%
\end{equation}
for $\varrho^{\alpha\overline{\beta}}$ a $(1,1)$ bivector compatible with the
symplectic structure induced by the K\"{a}hler form. The Moyal product for two
functions $f$ and $g$ is, to leading order in $\varrho$:%
\begin{equation}
f\ast g=fg+\varrho^{\sigma\overline{\tau}}\left(  \partial_{\sigma}%
f\partial_{\overline{\tau}}g-\partial_{\overline{\tau}}f\partial_{\sigma
}g\right)
\end{equation}
where $\sigma$ and $\overline{\tau}$ are holomorphic and anti-holomorphic
tangent bundle indices, respectively. Given a non-commutative $U(1)$ gauge
field $\mathcal{A}_{\overline{\alpha}}$ and a field $Q$ in the adjoint, the
covariant derivative is:%
\begin{equation}
\nabla_{\overline{\alpha}}Q=\partial_{\overline{\alpha}}Q+\mathcal{A}%
_{\overline{\alpha}}\ast Q-Q\ast\mathcal{A}_{\overline{\alpha}}.
\end{equation}
In terms of the Moyal expansion, we have:%
\begin{equation}
\nabla_{\overline{\alpha}}Q=\partial_{\overline{\alpha}}Q+\widetilde
{a}_{\overline{\alpha}}^{\overline{\tau}}\partial_{\overline{\tau}%
}Q-a_{\overline{\alpha}}^{\sigma}\partial_{\sigma}Q
\end{equation}
where we have introduced the modes:%
\begin{equation}
a_{\overline{\alpha}}^{\sigma}=2\varrho^{\sigma\overline{\tau}}\partial
_{\overline{\tau}}\mathcal{A}_{\overline{\alpha}}\text{, \ \ \ \ }\overline
{a}\overline{_{\overline{\alpha}}^{\tau}}=2\varrho^{\sigma\overline{\tau}%
}\partial_{\sigma}\mathcal{A}_{\overline{\alpha}}\label{gaugemodes}%
\end{equation}
Note that what actually couples to the mode $Q$ is the gradient of
$\mathcal{A}$ rather than $\mathcal{A}$ itself. This again is an indication
that the adjoint $U(1)$ behaves differently from other types of modes in the
fuzzy system.

The modes $a$ and $\overline{a}$ transform as $(0,1)$ forms valued
respectively in $T^{(1,0)}$ and $T^{(0,1)}$, i.e. the holomorphic and
anti-holomorphic parts of the tangent bundle. A zero mode of $Q$ is specified
to zeroth order by the condition $\partial_{\overline{\alpha}}Q=0$. To first
order, this is corrected to the condition $\partial_{\overline{\alpha}%
}Q-a_{\overline{\alpha}}^{\sigma}\partial_{\sigma}Q=0$. By inspection,
$a_{\overline{\alpha}}^{\sigma}$ defines an almost complex structure, and the
derivative $\nabla_{\overline{\alpha}}$ corresponds to the gauge field
associated with complex structure deformations. This is of course in keeping
with the expectation that twistor space is sensitive to complex structure
deformations, and so a putative bulk gravitational theory will be of
Kodaira-Spencer type.

To further bolster this interpretation, we can also work out the $(0,2)$
component of the field strength $\mathcal{F}_{\overline{\alpha}\overline
{\beta}}=\left[  \nabla_{\overline{\alpha}},\nabla_{\overline{\beta}}\right]
$. A lengthy but straightforward computation shows that the field strength
takes the form:%
\begin{equation}
\mathcal{F}=\overline{\partial}\mathfrak{A}+\left\{  \mathfrak{A}%
,\mathfrak{A}\right\}  _{\mathcal{L}}\label{FIELDSTRENGTH}%
\end{equation}
where $\mathfrak{A}=\frac{1}{2}\left(  a_{\overline{\alpha}}^{\sigma}%
\partial_{\sigma}+\overline{a}\overline{_{\overline{\alpha}}^{\tau}}%
\partial_{\overline{\tau}}\right)  $ is a $(0,1)$ form valued in the tangent
bundle of twistor space. Here, $\left\{  \mathfrak{A},\mathfrak{A}\right\}
_{\mathcal{L}}$ denotes the Lie bracket for the $(0,1)$-form $\mathfrak{A}$
taking values in the tangent bundle. In other words, the adjoint $U(1)$ of the
fuzzy system is naturally associated with the algebra of vector fields on
twistor space.

Given that the effective low energy theory describes physics derived from a
holomorphic Chern-Simons field theory, it is natural to expect that the geometric
action for the adjoint $U(1)$ gauge field $\mathfrak{A}$ can be packaged
in the form of a $BF$ type action that imposes the equation of motion
$\mathcal{F}^{(0,2)} = 0$. So it is reasonable to write an action of the form
\begin{equation}
S_{u(1)}=\int\Omega\wedge\widetilde{\mathfrak{A}}\wedge\left(  \overline
{\partial}\mathfrak{A}+\left\{  \mathfrak{A},\mathfrak{A}\right\}
_{\mathcal{L}}\right)  \label{SUU}%
\end{equation}
where $\widetilde{\mathfrak{A}}$ is a $(0,1)$-form in the cotangent bundle
which forms a pairing with the \textquotedblleft color
indices\textquotedblright\ of $\mathcal{F}$. The critical points of this
action enforce the condition that the
$(0,2)$ component of the field strength is compatible with the complex
structure of the geometry. Returning to our discussion between equations
(\ref{gaugemodes})-(\ref{FIELDSTRENGTH}), note further that to leading order
in the gauge field fluctuations, only the almost complex structure deformation
$a_{\overline{\alpha}}^{\sigma}$ appears in a zero mode equation. Setting
$\widetilde{a}\overline{_{\overline{\alpha}}^{\tau}}=0$, the field strength
$\mathcal{F}$ reduces to the Nijenhuis tensor for the almost complex structure.
Vanishing of the field strength enforces the condition that it is integrable.
This is of course quite familiar from earlier work on twistor string theory,
and the appearance of a Kodaira-Spencer theory of gravity, as in
\cite{Berkovits:2004jj}. BF type actions for the self-dual sector of supergravity have been
considered in \cite{Mason:2007ct, Mason:2008jy} (see also \cite{AbouZeid:2006wu}).

But as opposed to the twistor string, the coupling of the mode $\mathfrak{A}$
to the defect system comes with additional structure, as reflected in the
computation of MHV\ graviton amplitudes. Indeed, in the MHV amplitude, the
$gl(k_{N})$ charges which couple to $a_{\overline{\beta}}^{\sigma}%
\in\Omega^{(0,1)}(T^{(1,0)})$ are invariant under longitudinal shifts $v_{\dot a a} \to v_{\dot a a} + p_{\dot a a}$
in the polarization tensor, which allows one to pick the gauge condition $p\cdot v=0$.
The positive helicity gravitons then describe special modes for which the
$(0,1)$-form $a_{\overline{\beta}}^{\sigma}$   is Hamiltonian with respect to the infinity
twistor $I^{\alpha\beta}$ \cite{Mason:2008jy}:%
\begin{equation}
a_{\overline{\beta}}^{\sigma}=I^{\alpha\sigma}\partial_{\alpha}h_{\overline
{\beta}}.
\end{equation}
for $h_{\overline{\beta}}$ a line bundle valued $(0,1)$-form. A similar
redundancy condition holds for the dual mode $\widetilde{\mathfrak{A}}$
\cite{Mason:2008jy}, and the corresponding physical mode $\widetilde{h}$.
These have the interpretation as the two helicities of a graviton.

To give a bit more detail, recall that Einstein gravitons are given as elements
$h_{--}\in H^{1}(O(2))$ and $h_{++}\in H^{1}(O(-6))$, on $\mathbb{PT}^{\prime
}$, twistor space with a line at infinity removed. By contrast, the conformal
graviton is instead represented by $(0,1)$-forms in the holomorphic tangent
and cotangent bundle, i.e. $\mathcal{C}\in H^{1}(T^{(1,0)})$ and
$\widetilde{\mathcal{C}}\in H^{1}(\Omega^{(1,0)})$. With no additional
restrictions, $\mathcal{C}$ is simply the mode $a$ of line (\ref{gaugemodes}).
This is a linearized complex structure deformation for a Kodaira-Spencer
theory of gravity on twistor space. To get back the usual modes of Einstein
gravity, one introduces a redundancy for the $\mathcal{C}$ and $\widetilde
{\mathcal{C}}$ modes, so that only $h_{--}$ and $h_{++}$ survive as physical
excitations. As explained in \cite{Mason:2008jy}, the precise link between the
two is given by:%
\[
\mathcal{C}=I^{\alpha\beta}\frac{\partial h_{--}}{\partial Z^{\alpha}}%
\frac{\partial}{\partial Z^{\beta}}\text{, }h_{++}=I^{\alpha\beta}%
\frac{\partial\widetilde{\mathcal{C}}_{\beta}}{\partial Z^{\alpha}}%
\]
where $I^{\alpha\beta}$ is the inverse bitwistor for Minkowski space. The
redundancy is manifest for $\mathcal{C}$, and for $\widetilde{\mathcal{C}}$ is
obtained through the conditions:%
\begin{align}
\widetilde{\mathcal{C}} &  \rightarrow\widetilde{\mathcal{C}}+\partial
m+n\left\langle \pi d\pi\right\rangle \\
\widetilde{\mathcal{C}} &  \rightarrow\widetilde{\mathcal{C}}+\overline
{\partial}\chi
\end{align}
where $m$ is an element of $\Omega^{(0,1)}(O(-4))$ and $n$ is an element of
$\Omega^{(0,1)}(O(-6))$. In addition, $\chi$ is an element of $\Omega
^{(1,0)}(O(-4))$, and the corresponding redundancy equation indicates
$\widetilde{\mathcal{C}}$ is represented by a cohomology class.


On the 4D space-time, the exchange of the $\widetilde{h}$/$h$ pair between
twistor lines (i.e. space-time points) then generates the usual gravitational
potential for a spin two excitation with the strength of Newton's constant set
by the ratio of an overall coefficient multiplying (\ref{SUU}) and the bulk
defect coupling (\ref{defectcoup}). It would be interesting to refine this discussion further
and fix the precise value of Newton's constant.

\subsection{Einstein versus Conformal Gravity \label{ssec:CONF}}

What is the low energy effective action for this putative theory of gravity?
The leading order terms consistent with the symmetries of the system are:
\begin{align}
S_{grav}=\alpha\Lambda\int d^{4}x\sqrt{-g}\left(  R-2\Lambda\right)
+\beta\int d^{4}x\sqrt{-g}W^{2}+...
\end{align}
where $R$ is the scalar curvature, $\Lambda$ is the cosmological constant, $W$
is the Weyl tensor, and $\alpha$ and $\beta$ are dimensionless couplings. In
Einstein gravity, we would set $\alpha=1/(16 \pi G_{N}\Lambda)$. The case
$\alpha=0$ corresponds to conformal gravity. This latter theory contains a
number of pathologies compared to ordinary Einstein gravity. For example, the
kinetic term is quartic in the momentum, and the resulting spectrum of modes
contains ghost-like excitations. This is also the
gravitational sector of the twistor string theories of \cite{Witten:2003nn,
Berkovits:2004hg, Berkovits:2004jj}.

We will now list a set of arguments that indicate that in our case, rather than getting
conformal gravity, the emergent gravity theory is described by the Einstein action with a
Newton coupling set by $\sqrt{G_{N}} \sim \ell_{pl} \sim \ell/\sqrt{N}$.

$\bullet$ {\it Graviton scattering}: In principle, we can determine the effective action by
computing all possible graviton scattering amplitudes which descend from the matrix model.
So far, we have only computed the MHV amplitudes. Are these sufficient to distinguish the
two theories? This is in fact a somewhat subtle question, because some features of Einstein
gravity in (anti-)de Sitter space can be mimicked by conformal gravity
with Neumann boundary conditions at spatial infinity \cite{Maldacena:2011mk}.

MHV graviton scattering amplitudes describe the process of an incoming
negative helicity graviton bouncing off a self-dual space-time background.
Self-dual backgrounds are solutions to both the Einstein and the conformal theory.
So we can set up the calculation in both theories, by starting with a self-dual geometry
that asymptotically looks like an (anti-)de Sitter space-time. Conformal gravity contains
two propagating graviton modes, one that behaves like the ordinary graviton and
a ghost mode. By picking appropriate asymptotic conditions, we can isolate the
Einstein graviton. The conformal gravity calculation then looks identical to the
Einstein gravity calculation. The difference between the two theories is that
one theory has a dimensionful constant and the other a dimensionless one.
The reason that we can compare the two directly is because we are considering the
system on de Sitter space with finite curvature radius $\ell^2 \sim 1/\Lambda$.

For the sake of argument, let us suppose that the gravitational theory described by the
matrix model is dominated by the conformal gravity term proportional to $\beta$.
The effective Newton's constant would then fixed by the coefficient $\beta$ and the
de Sitter radius \cite{Maldacena:2011mk}:
\[
G_{N}\sim\frac{1}{\beta\Lambda}.
\]
If we were to match our space-time theory to this prescription, we would need
to set $\Lambda\sim1/\ellcc^{2}$ and $G_{N} \sim \ell_{pl}^{2}\sim\ellcc^{2}/N$
which would imply that $\beta\sim N$. In the large $N$/flat space scaling
limit, this looks absurd. Indeed, conformal gravity (with any finite coupling $\beta$)
does not have a non-zero MHV graviton scattering amplitude in flat space.\footnote{
Taking $\beta \sim N \to \infty$ amounts to taking  a classical limit in conformal gravity. One could suspect
that this classical limit amounts to a decoupling limit, in which the ghost
mode of conformal gravity can be consistently decoupled from the rest of the theory.
If this were possible, the left over graviton mode would propagate and interact like
an ordinary graviton in Einstein gravity, with a finite Planck length equal to $\ell_{pl}$.}
Further, explicit calculations of graviton scattering in twistor string theory vanish for
Einstein gravitons \cite{Dolan:2008gc}. Our double scaled matrix model does have finite
flat space scattering amplitudes.

\smallskip

$\bullet$ {\it Absence of conformal symmetry at finite $\ell_{pl}$}:  Conformal symmetry is explicitly broken
in the double scaled matrix model, both due to the presence of a UV cut-off $\ell_{pl}$ and
due to the introduction of the infinity bi-twistor. In the non-commutative twistor theory,
only the compact $SU(4)$ subgroup of the complexified conformal group $SL(4,\mathbb{C})$ is unitarily realized.
Moreover, the defect action, which involves the infinity twistor,  further breaks the symmetry to
$SO(5)$. The $SO(5)$ generators therefore play a distinguished role among the
currents of the defect system. The (MHV manifestation) of the graviton modes
are made up from these $SO(5)$ currents -- specifically, those associated with the
hermitian `translation' generators (\ref{calp}). The hermitian generators (\ref{calp})
are precisely the ones that are compatible with the boundary conditions \cite{Maldacena:2011mk}
that selects the Einstein gravitons from the conformal gravity fluctuations.\footnote{Note also that,
by starting from the $S^4$ with finite radius $\ell$, we are able to define hermitian translation generators
(\ref{calp}), and avoid the non-unitarity that plagued the twistor string theory of \cite{Berkovits:2004jj}.}

\smallskip

$\bullet$ {\it CSW correspondence:} As explained in section 6.1 and in \cite{GMM}, integrating
out the defect modes produces an effective action that, in the continuum limit, matches with
the MHV effective action in \cite{Boels:2006ir, Boels:2007qn}, which (by design in
\cite{Boels:2006ir, Boels:2007qn}, but derived here) evades the troubling appearance of
conformal gravity amplitudes, that plague the twistor string proposal of \cite{Witten:2003nn, Berkovits:2004jj}.
The appearance of gravity for us indeed has a different geometric origin, in which the
Einstein gravitons immediately play a distinguished role.

\smallskip

Suppose the theory does indeed give rise to Einstein gravity, as we have argued.
What kind of theory is it, and does the matrix model -- or rather its UV
parent theory, the holomorphic Chern-Simons theory with flux -- provide a UV
complete description? With regard to the first question, indications are that
the IR theory will take the form of ${\cal N}=4$ supergravity, possibly coupled
to ${\cal N}=4$ gauge theory. We have not found a clear condition that limits
the possible rank of the gauge group. This suggests that the answer to the
second question should be: No, the matrix model (and even its embedding in
the topological B-model string theory) is not UV complete, but should be
viewed as the low energy effective description of a more complete theory.
To illustrate how such an embedding may arise, we now describe
a potential realization of the matrix model in the physical superstring.

\section{Embedding in Superstring Theory \label{Embedder}}

In this section we propose an embedding of the twistor matrix model in a
brane construction in superstring theory. In particular, this will establish a
concrete UV\ completion for our system. The basic setup we consider is given
by a D0-brane bound to a stack of eight-branes in type IIA\ superstring
theory. We work in ten non-compact space-time dimensions, which we write as
$\mathbb{R}_{time}\times\mathbb{R}^{8}\times\mathbb{R}_{\bot}$. The
eight-branes are assumed to be parallel, filling $\mathbb{R}_{time}%
\times\mathbb{R}^{8}$ and sitting at various points of $\mathbb{R}_{\bot}$.
The D0-brane we shall be considering will be bound to one stack of
eight-branes, but will be free to move in the worldvolume $\mathbb{R}%
_{time}\times\mathbb{R}^{8}$. The basic idea is that when a suitable flux is
switched on along the worldvolume of the eight-branes, we obtain four copies
of the supersymmetric harmonic oscillator. The statistical mechanics of this
system is studied by compactifying on a thermal circle of radius $\beta$. We
argue that the many body problem for a gas of 0-8 strings at low temperature
realizes the twistor matrix model.

\subsection{Brane Construction}

Our main interest will be in the worldvolume theory of a single D0-brane
probing a stack of $N_{D8}$ D8-branes coincident with an O8-plane. However, to
frame our discussion, let us first briefly review the worldvolume theory of
$N_{D0}$ D0-branes, first in the absence of other branes, and then in the
presence of eight-branes.

Consider first the case of $N_{D0}$ D0-branes in ten flat space-time
directions. This is a system which preserves sixteen real supercharges, and is
described at low velocity by a quantum mechanics with $(8,8)$ supersymmetry.
For $N_{D0}$ D0-branes, we have a $U(N_{D0})$ gauge theory with gauge field
$A_{D0}$. All of the 0-0 strings transform in the adjoint of $U(N_{D0})$. The
bosonic matter content consists of eight real scalars $X^{I}$ for $I=1,..,8$
which describe motion parallel to the D8-brane and an additional real scalar
$X_{\bot}$ describing motion transverse to the D8-brane. In addition there are
sixteen fermions, which transform as a Majorana-Weyl spinor of $so(9,1)$.
Under the decomposition $so(9,1)\supset so(8)\times so(1,1)$, these fermions
transform in the $8_{s}$ and $8_{c}$ of $so(8)$. We denote these two sets by
$S_{A}\oplus\widetilde{S}_{A^{\prime}}$ for $A,A^{\prime}=1,...,8$.

Introducing a stack of D8-branes leads to additional dynamics for the system.
Due to their high codimension, adding D8-branes introduces a number of
subtleties. For example, at finite distance from the D8-brane, the dilaton
blows up. Consistent treatment of the system then requires introducing
O-planes to prevent this.\ When the tadpole is not locally cancelled, there
will also be a Romans mass on one side of the D8-brane
\cite{Polchinski:1995df}.

The effective theory of the D0-brane in the presence of $N_{D8}$ D8-branes has
been treated previously in the literature, see for example
\cite{Bachas:1997kn}. This system preserves a chiral $(8,0)$ supersymmetry. In
addition to the 0-0 strings, the worldvolume theory now contains an additional
0-8 string charged in the $(N_{D8},\overline{N_{D0}})$, which is a complex
fermion we denote by $\chi$. This mode is a singlet under the (on-shell)
supersymmetry transformations, which is possible due to the low dimension and
chiral nature of the supersymmetry. The resulting D0-D8 system is BPS. In
particular, the D0-brane experiences no force in the direction transverse to
the D8-brane. See \cite{Bachas:1997kn} for further discussion.

The system we shall mainly be interested in for our purposes is the case of
$N_{D8}$ D8-branes coincident with an O8-plane. Local tadpole cancellation
then fixes $N_{D8}=16$, that is, the maximal gauge group is $SO(16)$. Let us
note that $E_{8}$ can also be realized, though non-perturbatively. The
presence of the orientifold plane affects the 0-0 and 0-8 strings as well. The
D0-brane gauge group is now $O(N_{D0})$.\footnote{As explained for example in
\cite{Schwarz:1999vu} the reason the gauge group is $O(N_{D0})$ rather than
$SO(N_{D0})$ is because upon compactifying on the transverse circle $S_{\bot
}^{1}$, there are $\mathbb{Z}_{2}$ valued Wilson line configurations available
in the T-dual description.} The 0-0 strings, which were initially in the
adjoint of $U(N_{D0})$ are now two index representations of $O(N_{D0})$. The
modes in the two index symmetric representation descend from $X^{I}\oplus
S_{A}$, while the modes in the two index anti-symmetric descend from
$A_{D0}\oplus\widetilde{S}_{A^{\prime}}\oplus X_{\bot}$. The 0-8 strings now
transform in the $(N_{D0},N_{D8})$, which is a real representation. To
maintain a holomorphy convention, we shall often decompose $SO(N_{D8})\supset
U(N_{D8}/2)$. The 0-8 string content then consists of a vector-like pair of
complex fermions $\chi\oplus\widetilde{\chi}$, which are subject to the
reality condition $\chi^{\dag}=\widetilde{\chi}$.

Consider now the special case of a single D0-brane. When $N_{D0}=1$, we
project out the mode describing motion transverse to the D8 stack, and also
eliminate half of the fermionic modes. The gauge group is also in this case
reduced to $O(1)=%
\mathbb{Z}
_{2}$. The effective Lagrangian is (in units where the D0-brane mass is
unity):
\bea
L_{eff}=\frac{1}{2}\partial_{t}X^{I}\partial_{t}X^{I}+\frac{i}{2}S_{A}%
\partial_{t}S_{A}+\frac{i}{2}\left(  \widetilde{\chi}D\chi+\chi D\widetilde
{\chi}\right)
\eea
where the covariant derivative acts on the fermionic modes as:%
\begin{align}
D\chi &  =\left(  \partial_{t}+igA_{D8}\right)  \chi\\
D\widetilde{\chi}  &  =\left(  \partial_{t}-igA_{D8}\right)  \widetilde{\chi}%
\end{align}
where $A_{D8}$ is the pullback of the D8-brane gauge field to the D0-brane
worldvolume, with $g$ the gauge coupling of the D8-brane theory. In the above,
we have specialized to the case of a gauge field activated in the
$U(N_{D8}/2)$ subgroup.

This system describes a D0-brane which is free to move inside of the D8-brane.
In the quantum theory, the conjugate momenta to $X^{I}$ are $\Pi^{I}%
=\partial_{t}X^{I}$. This results in the commutation relations:%
\bea
\left[  X^{I},\Pi^{J}\right]  =i\delta^{IJ}\text{, }\left\{  S_{A}%
,S_{B}\right\}  =\delta_{AB}\text{, }\left\{  \chi_{k},\chi_{\overline{l}%
}^{\dag}\right\}  =\delta_{k\overline{l}}%
\eea
where $k$ (resp. $\overline{l}$) denotes an index in the fundamental (resp.
anti-fundamental) of $U(N_{D8}/2)$. Observe that there is a degeneracy of
ground states for the system. These correspond to the possible positions of
the D0-brane in the superspace $\mathbb{R}^{8|8}$. We will soon switch on a
flux which identifies a canonical choice of complex structure, in which case
we have $\mathbb{C}^{4|4}$.

To break this degeneracy in the ground state energies we now switch on a
magnetic flux which takes values in the overall $U(1)\subset U(N_{D8}/2)$. The
effective energy scale for the flux is a tunable parameter of the D8-brane
worldvolume theory. Since the D0-brane is more accurately thought of as a
superparticle, it is appropriate to consider a background gauge field
configuration:%
\bea
A_{D8}=F_{IJ}^{(v)}X^{I}\partial_{t}X^{J}+F_{AB}^{(s)}S^{A}S^{B}%
\eea
where here, $F^{(v)}$ and $F^{(s)}$ reflect the presentation of the flux in
the $8_{v}$ and $8_{s}$ of $so(8)$. By triality, we can take an embedding of
$su(4)\times u(1)\subset so(8)$ so that the $8_{v}$ and $8_{s}$ decompose as:%
\begin{align}
so(8)  &  \supset su(4)\times u(1)\\
8_{v}  &  \rightarrow4_{+1}+\overline{4}_{-1}\\
8_{s}  &  \rightarrow4_{-1}+\overline{4}_{+1}%
\end{align}
The original basis of modes can now be traded for the usual coordinates of
supertwistor space, $Z^{\alpha}\oplus\psi^{i}$. In terms of this basis, the
pullback of the gauge field is:%
\bea
A_{D8}=F_{\alpha\overline{\beta}}\left(  Z^{\alpha}\partial_{t}\overline
{Z^{\beta}}-\overline{Z^{\alpha}}\partial_{t}Z^{\beta}\right)  +F_{i\overline
{j}}\psi^{i}\overline{\psi^{j}}%
\eea
The profile of this flux is taken to be:%
\bea
F=i\frac{f_{1}}{2}dz^{1}\wedge d\overline{z^{1}}+i\frac{f_{2}}{2}dz^{2}\wedge
d\overline{z^{2}}+i\frac{f_{3}}{2}dz^{3}\wedge d\overline{z^{3}}+i\frac{f_{4}%
}{2}dz^{4}\wedge d\overline{z^{4}}.
\eea
In general, this choice of flux will break supersymmetry on the D8-brane,
though the D0-brane may still experience an effective supersymmetric quantum
mechanics.\footnote{Following \cite{Mihailescu:2000dn, Witten:2000mf, Fujii:2001wp}, to study the unbroken
supersymmetries in the presence of this flux, introduce the block diagonal
matrix $M$ in the $8_{v}$ representation (see e.g. \cite{Fradkin:1985qd,
Abouelsaood:1986gd, Callan:1986bc, Witten:2000mf}):%
\bea
M=\underset{i=1}{\overset{4}{\oplus}}M_{i}\text{, }M_{i}=\left[
\begin{array}
[c]{cc}%
\cos2\pi v_{i} & \sin2\pi v_{i}\\
-\sin2\pi v_{i} & \cos2\pi v_{i}%
\end{array}
\right]  .
\eea
where each $v_{j}$ is given by:%
\bea
\exp2\pi iv_{j}=\frac{1+if_{j}}{1-if_{j}}.
\eea
The condition for unbroken supersymmetry is that in the $8_{c}$
representation, $\rho(-M)$ has a unit eigenvalue. The amount of supersymmetry
which is preserved is dictated by the maximal subalgebra of $u(4)\subset
so(8)$ which commutes with $\rho(-M)$. \ A necessary condition for some
supersymmetry to be preserved is:%
\bea
v_{1}+...+v_{4}\in%
\mathbb{Z}
.
\eea
For a generic choice which retains some supersymmetry, this leaves a system
with only two real supercharges. The case we shall eventually specialize to is
given by taking all $f_{j}=f$. In this case, we see that generically, all
supersymmetry will be broken. However, in the limit $f\rightarrow+\infty$, we
see that all eight supercharges are retained. This corresponds to the limit
$v_{i}\rightarrow1/2$.}

Let us now study the quantum mechanics of the endpoint for the 0-8 string
which is attached to the D8-brane. Evaluating in the background where the 0-8 string
is present, the Lagrangian is:%
\bea
L_{eff}=\frac{1}{2}\partial_{t}X^{I}\partial_{t}X^{I}+\frac{i}{2}S_{A}%
\partial_{t}S_{A}+igF_{IJ}^{(v)}X^{I}\partial_{t}X^{J}+igF_{AB}^{(s)}%
S^{A}S^{B}.
\eea
Quantization is now of the standard type for a superparticle moving in a
background magnetic flux. This is described by four supersymmetric harmonic
oscillators, which by abuse of notation we label as $Z^{\alpha}$ and $\psi
^{i}$. These modes have the same supercommutators as fuzzy supertwistor space,
with characteristic frequencies:%
\bea
\omega_{\alpha}=\frac{2f_{\alpha}}{m_{D0}}\text{.}%
\eea
Here we have rescaled the coordinates to include the explicit dependence on
the D0-brane mass. The effective Hamiltonian is:%
\bea
H_{eff}=\underset{\alpha=1}{\overset{4}{\sum}}\omega_{\alpha}Z_{\alpha}^{\dag
}Z^{\alpha}+\underset{i=1}{\overset{4}{\sum}}\omega_{i}\psi_{i}^{\dag}\psi^{i}%
\eea
where the zero point energy cancels out. In what follows we focus on the
special case of $\omega=\omega_{\alpha}=\omega_{i}$ for all $\alpha$ and $i$.
Indeed, in this case is proportional to the Hamiltonian constraint operator
$H_0$ of equation (\ref{superconstraint}).

An important feature of this system is that it admits a decoupling limit,
where we take $f$ to be large in string units, while keeping the ratio
$f/m_{D0}$ small in string units. This is basically the same type of limit
studied in \cite{SeibergWitten} for non-commutative Yang-Mills theory. Here,
this zero slope limit ensures that the low energy dynamics are well-described
by this effective Hamiltonian.

\begin{figure}[ptb]
\begin{center}
\includegraphics[
height=2.2425in,
width=1.9657in
]{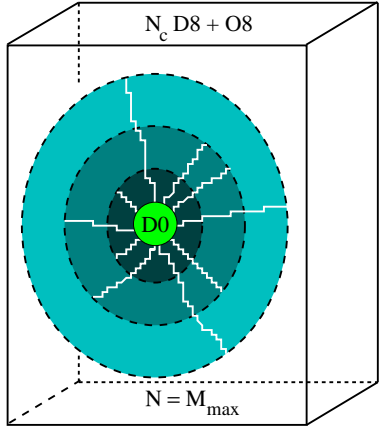}
\end{center}
\caption{Depiction of the D0-brane bound to a stack of $N_{c}$ D8-branes and
an O8-plane in the presence of a magnetic flux threading the D8-brane. The 0-8
strings are attached to the D0-brane, and the other end can wander inside the
D8-brane. Switching on a background flux through the D8-branes, a gas of 0-8
strings descend into Landau levels graded by integers $M\geq0$. Here, we have
depicted a low temperature configuration which realizes fuzzy supertwistor
space at level $N=M_{\max}$.}%
\label{d0d8levels}%
\end{figure}

The modes at level $M$ are described by the Hilbert space of fuzzy points for
$\mathbb{CP}^{3|4}$, $\mathcal{H}_{\mathbb{CP}^{3|4}}(M)$, so there is a
degeneracy of $K_{M}$ at each $M$. Thus, the phase space available to a 0-8
string is now a direct sum of fuzzy supertwistor spaces. Finally, at each
level, there is an additional overall degeneracy, because the fermionic 0-8
string transforms in the fundamental of $U(N_{D8}/2)$. We denote the effective
$N_{c}=N_{D8}/2$. See figure \ref{d0d8levels} for a depiction of this system.

\subsection{Statistical Mechanics}

Let us now study the statistical mechanics of a large number of 0-8 strings.
We compactify the temporal direction on a circle of circumference $\beta=1/T$
and take thermal boundary conditions for all modes. Though there are some
subtleties in such stringy systems at high temperature, i.e. small thermal
radius, we shall mainly confine our discussion to the low temperature limit.

In the grand canonical ensemble, the chemical potential for the 0-8 strings is
controlled by a flat connection from the D8-brane gauge field. The full gauge
field configuration on the D8-brane is then given by the magnetic flux
considered earlier, and a holonomy from the component which points along the
thermal circle direction. The energetics of the 0-8 strings are unaffected by
including the contribution from the flat connection, though it does introduce
a chemical potential for the system. Integrating over the thermal circle, we
obtain the fugacity $\zeta$ for the grand canonical ensemble:%
\bea
\zeta=\exp\left(  \int_{S_{\beta}^{1}}A_{D8}\right)  .
\eea
The chemical potential is then given by $\mu=T\log\zeta$.

Consider first the zero temperature limit, with a fixed number of 0-8 strings
which we denote by $N_{tot}$. The phase space for each particle is a
non-commutative $\mathbb{C}^{4|4}$. Since the 0-8 strings are fermionic, they
will start to fill up a Fermi surface. We will mainly be interested in the
case where the Fermi surface forms a complete shell up to a maximal level
$M_{\max}=N$. This corresponds to taking a fixed number of states:%
\bea
N_{tot}=\underset{M=0}{\overset{M_{\max}}{\sum}}K_{M}\text{.}%
\eea
Each energy shell corresponds to the Hilbert space of a fuzzy $\mathbb{CP}%
^{3|4}$ at level $M$. As we increase $M_{\max}$, we approach a large $N$
limit. The states fill out entries in the direct sum:%
\bea
\mathcal{H}(N_{tot})=\mathcal{H}_{\mathbb{B}^{(4|4)}}(N)\otimes N_{c}%
=\underset{M=0}{\overset{M_{\max}}{\oplus}}\mathcal{H}_{\mathbb{CP}^{3|4}%
}(M)\otimes N_{c}%
\eea
where the tensor product with the factor of $N_{c}$ reflects the fact that
each particle transforms in the fundamental of $U(N_{c})$. Up to this factor,
we can identify this Hilbert space with the fuzzy four-ball encountered
previously in section \ref{sec:TMM}.

The length scale $\ellcc$ is set by the inverse characteristic frequency of
the system $\omega$:%
\bea
\ellcc\sim\omega^{-1}\text{.}%
\eea
We can also introduce an anisotropic scaling for the D8-brane magnetic flux,
and thereby include the parameter $\gamma$ of the infinity bitwistor.

Returning to the case of uniform energy level spacing in the phase space, we
would now like to model the effective dynamics of the many body system at
finite temperature. It is helpful to work in terms of fields, which describe
the collective excitations. There are two types of excitations which we can
identify, corresponding to motion within the Fermi surface, and normal to it.

Excitations above the Fermi surface are 0-8 strings, which are characterized
by a field $Q(Z^{\dag},\psi^{\dag},Z,\psi)$ in the fundamental of $U(N_{c})$.
Holes are 8-0 strings, and are characterized by a field $\widetilde{Q}%
(Z^{\dag},\psi^{\dag},Z,\psi)$ in the anti-fundamental. Moving the location of
the D0-brane, we see that there are also four complex scalars. These can be
viewed as bound states of the $8-0\oplus0-8$ strings. If the particles were
free to move inside of the D8-brane, we would have four such complex scalars
$\mathcal{A}^{\alpha}(Z^{\dag},\psi^{\dag},Z,\psi)$.\footnote{Although it is
tempting to also include four complex fermionic modes $\Psi^{i}(Z^{\dag}%
,\psi^{\dag},Z,\psi)$, the geometric interpretation of these modes is less
clear, as we cannot \textquotedblleft give a vev\textquotedblright\ to a
fermionic direction.} These modes define the transformation of the states
within the Fermi surface, and are therefore maps $\mathcal{H}(N_{tot}%
)\rightarrow\mathcal{H}(N_{tot}+1)$. In other words, they are fuzzy
$(0,1)$-forms acting on $\mathcal{H}_{\mathbb{B}^{(4|4)}}(N)$ which are also
in the adjoint representation of $U(N_{c})$. Since we are interested in the
low energy dynamics confined to near the Fermi surface, there is an additional constraint:
\begin{equation}
Z_{\alpha}^{\dag} \mathcal{A}^{\alpha} = 0.
\end{equation}
so there are only three independent bosonic degrees of freedom.

We fix the effective action via symmetry considerations. The kinetic operator
for the particle/hole pairs corresponds to an excitation in the direction
normal to the Fermi surface. In a local patch where we write $\mathbb{CP}%
^{3|4}\approx\mathbb{C}^{2|4}\times\mathbb{CP}^{1|0}$, this is in the
direction of the bosonic twistor line. From our earlier analysis of the defect
kinetic operator, we know that a derivative operator which is local on the
\textquotedblleft space-time\textquotedblright\ directions of the twistor
fibration is the kinetic operator $\overline{\mathcal{D}}$. Symmetry
consideration then dictate the kinetic operator for the particle/hole pair at
low temperatures:%
\bea
S_{\text{defect}}=\text{Tr}_{\mathcal{H}(N_{tot})}\left(  \mathcal{I}_{IJ}\widetilde
{Q}\mathcal{D}^{I}Q\Omega^{J}\right)
\eea
where the form of the derivative operators is fixed by gauge invariance. Here,
the defect action is integrated over the entire four-ball of the phase space.
Excitations near the top of the Fermi surface cost less energy, and so at low
energies, we get back the integral over just the top layer, given by the
$\mathbb{CP}^{3|4}$ at level $N=M_{\max}$.

For the bulk modes, there is a single topological term we can write down which
is consistent with the symmetries of the system; it is the holomorphic volume
of the phase space in the supermanifold $\mathbb{C}^{4|4}$ occupied by the 0-8
strings:%
\bea
S_{\text{bulk}}\left(  \beta\right)  =\int_{S_{\beta}^{1}}C_{1}\wedge
\text{Tr}_{\mathcal{H}(N_{tot})}\left(  \varepsilon_{\alpha\beta\gamma\delta
}D^{\alpha}D^{\beta}D^{\gamma}D^{\delta}\right)  _{\psi^{4}}%
\eea
where $D^{\alpha}=Z^{\alpha}+\mathcal{A}^{\alpha}$ and $C_{1}$ is the RR
one-form potential. The coupling constant of the matrix model is obtained by
integrating over the Euclidean thermal circle. Taking $C_{1}$ to be a constant
one-form and integrating over the circle yields an overall coupling.
In other words, the action becomes:
\bea
S_{\text{bulk}}\left(  \beta\right)  = \frac{1}{g^{2}}\text{Tr}%
_{\mathcal{H}(N_{tot})}\left(  \varepsilon_{\alpha\beta\gamma\delta}D^{\alpha
}D^{\beta}D^{\gamma}D^{\delta}\right)  _{\psi^{4}}%
\eea
Hence, the low temperature dynamics of a large number of 0-8 strings
reproduces the basic form of the twistor matrix model.

By embedding the matrix model in the physical superstring, we have equipped
the model with a proposed UV completion. Consistency of the UV completion leads to
some interesting restrictions on the form of the low energy effective theory.
For example, it imposes a sharp upper bound on the rank of the gauge group.
Note, however, that this restriction is milder than what is required in
conformal $\mathcal{N}=4$ supergravity for anomaly cancellation (see
\cite{Fradkin:1985am} for a review). A further interesting feature of this
construction is that it is formulated with ten non-compact directions. Indeed,
the embedding of the twistor matrix model into the superstring does not
involve a traditional compactification at all.\footnote{In string theory the
usual situation is that without a compactification of the \textquotedblleft
extra dimensions\textquotedblright\ one obtains a 4D gauge theory which is
\textit{decoupled} from gravity. For example, a D3-brane probing a non-compact
Ricci flat manifold will achieve this. Here, the interpretation of the ten
dimensions for the 4D theory is different, and gravity emerges along with the
space-time.}

Using this perspective, we can ask how various features of the 4D space-time
are encoded in the gas of 0-8 strings. First note that the double scaling
limit for realizing a flat space theory is quite natural in this setup; we can
scale $N\rightarrow\infty$, but simultaneously decrease the energy spacing
between the Landau levels. The currents of the matrix model are bilinears in
the $Q\widetilde{Q}$ modes, corresponding to the creation of particle/hole
pairs. If we are at low energies, all of these states will be created near the
top of the Fermi surface, and at the same Landau level. However, as we
increase the available energy, holes can be created further down in the Fermi
sea, and can also jump between neighboring Landau levels. Thus, a continuous
extra dimension opens up in the direction normal to the Fermi surface. It is
tempting to identify this with a holographic RG scale. At even higher
excitation energies, the particles are no longer close to the Fermi surface,
and simply move throughout the fuzzy $\mathbb{C}^{4|4}$ phase space. Note that
even in this case, however, there is an upper bound on the available
excitation energies.

Though at finite $N$ the continuum theory interpretation is less evident, it
is also the case where we have a truncated Hilbert space for Euclidean de
Sitter. In this case, the jump in Landau levels is discontinuous. A further
curious feature is that the effective ratio $\ellcc/\ell_{pl}\sim\sqrt{N}$
is modified at high energies so that states would appear to experience
distinct values of the cosmological radius. One can also consider the high
temperature regime -- i.e. high compared to the level spacing -- of the gas of
0-8 strings, but below the Hagedorn temperature. This is still a well-defined
system, but now the Fermi surface itself will begin to break apart due to
thermal fluctuations. Clearly, it would be interesting to study this set of
issues further.

\subsection{Dual Descriptions}

The D0-D8 bound state we have been considering has a number of dual
formulations, which provide additional insight into the dynamics of the
system. Compactifying on a circle $S_{\bot}^{1}$ in the direction transverse
to the eight-branes, we obtain a T-dual description in type I\ string theory,
where the D0-brane is now a D1-brane. This is in turn dual to an F1-string of
IIB via an S-duality. Using type I/heterotic duality, we can map this to a
heterotic string wrapping $\mathbb{R}_{time} \times S_{\bot}^{1}$. Observe
that the projection on the mode content we have observed is quite natural on
the heterotic string side. There is also an intriguing connection to the
$\mathcal{N}=2$ string. Indeed, as advocated for example in \cite{VafaFTHEORY}%
, the worldvolume theory of the D1-brane is naturally viewed as a
four-dimensional theory on $\mathbb{R}^{2,2}$ which has been reduced along a
null vector. This suggests a more direct connection between $\mathcal{N}=2$
strings and the physics of supertwistor space which would be interesting to
develop further.

We can also use these dual formulations to realize more intricate gauge
theories. The basic idea is to consider the low energy quantum mechanics of a
D0-brane, now bound to a more general configuration of eight-branes and lower
dimensional branes. Compactifying on a circle, this type of configuration is
T-dual to a configuration of intersecting seven-branes in IIB. Going to strong
coupling, we can study the resulting dynamics in terms of an effective
heterotic string probing an appropriate flux background.

As an example, consider three stacks of branes which we denote as
$\mathcal{B}$, $\mathcal{B}^{\prime}$ and $\mathcal{B}^{\prime\prime}$ which
wrap a common $\mathbb{R}_{time}\times\mathbb{R}^{4}$. We assume $\mathcal{B}$
describes a D8-brane, as in our previous discussion. Though it is beyond the
scope of our present discussion to provide a full description of this system,
we can sketch how additional matter sectors could arise. In addition
to the $0-\mathcal{B}$ strings, there will now be $0-\mathcal{B}$ and
$0-\mathcal{B}^{\prime\prime}$ strings. Excitations above the Fermi surface
can therefore be of different particle/hole types: They can correspond to
pairs of $\mathcal{B}-0\oplus0-\mathcal{B}$ modes, which we previously
identified with the modes of an $\mathcal{N}=4$ supermultiplet in the adjoint
of $U(N_{c})$. In addition, we see that there are new effective modes, such as
$\mathcal{B}-0\oplus0-\mathcal{B}^{\prime}$ which are in the fundamental of
the emergent $U(N_{c})$. Observe that gauge invariance also dictates the form
of the possible interaction terms of these composites.

Note also that the effective dynamics is not limited to unitary gauge groups.
Indeed, using the dual description in the heterotic string, there is an
effective worldsheet description available in the strong coupling limit of the
IIB\ configurations, so we can also consider E-type gauge theories. This
suggests that phenomenologically relevant gauge theories may also be
engineered in this type of setup (by embedding in a GUT group). It would be of
interest to provide a more explicit constructions along these lines.

\section{Conclusions \label{sec:CONC}}

In this paper we have provided evidence that $\mathcal{N}=4$ SYM\ theory on an
$S^{4}$ is dual to a large $N$ matrix model defined over non-commutative
twistor space. The matrix model is the theory of the lowest Landau level for
holomorphic Chern-Simons on $\mathbb{CP}^{3|4}$ expanded around the Yang
monopole configuration. The matrix model incorporates the symmetries of the
original space-time theory, and moreover, contains a natural class of spin 1
and spin 2 symmetry currents. Quite remarkably, in the flat space limit of the
matrix model, the correlators of these currents correctly reproduce MHV\ gluon
and graviton scattering \cite{GMM}. This is a highly non-trivial feature, and
provides evidence that the low energy limit is described by Einstein gravity.
We have also presented a proposed UV completion of the model based on the
low energy dynamics of a D0-D8 bound state.
In the remainder of this section we discuss various open directions.

At a pragmatic level, it is important to check that the match to the
4D theory continues to work at loop level in the gauge theory. Indeed, loop
level contributions from conformal graviton states is a significant hurdle for
the general twistor string program. This would also provide another probe into
the nature of the gravitational theory which is coupled to the gauge theory.

The most intriguing feature of the twistor matrix model is that it seems to
contain an emergent gravitational sector. Though there are strong constraints
on emergent theories of gravity, the basic setup of the matrix model violates
some assumptions of the Weinberg-Witten theorem \cite{Weinberg:1980kq}.
For example, the graviton and the space-time emerge simultaneously
in a double scaling limit. Furthermore, the
theory is defined as the infinite radius limit of a theory with positive
curvature. Additionally, it is far from evident that our model possesses
a local gauge invariant stress energy tensor.\footnote{One could ask whether the model comes with
additional higher spin currents, which did appear in \cite{Zhang:2001xs}. Here, note that
the interactions involving our spin 2 excitation are suppressed by the small parameter $\ell_{pl}$, and
so in the large $N$ limit, it is natural to expect any higher spin excitations to effectively decouple. Even
in the finite $N$ setting, these excitations are more appropriately viewed as
diffeomorphisms on the fuzzy twistor space, and so are not expected to be 4D higher spins.
It would nevertheless be interesting to study this issue further, and in particular, to see
whether a truncated version of a higher spin theory
along the lines of \cite{Vasiliev:1992av, Vasiliev:1995dn, Vasiliev:1999ba, Vasiliev:2003ev}
could be implemented in a variant of the matrix model considered in this paper.} Note, however, that although
there is a short distance cutoff $\ell_{pl}$, the low energy
theory will not contain Lorentz violating operators. This is simply because
the model respects $so(5)$ and its Wigner-In\"on\"u contraction in the flat space limit.
To further test our conjecture, it would be interesting to see
whether other features of quantum gravity such as black hole formation,
or a more direct account of holographic entropy bounds are also realized.

We have also seen that fuzzy twistor geometry and the matrix model may
naturally arise from a D0-D8 bound state. It is likely that this construction
generalizes to situations with less supersymmetry and more realistic matter
content. Observe that all ten of the string space-time dimensions are necessary
to realize the 4D dual theory. Thus, four space-time dimensions is a rather
special part of the construction.

Finally, as the 4D theory is formulated on a four-sphere, it is natural to
analytically continue the theory to de Sitter space. In particular, the large
but finite $N$ version of the matrix model suggests an intrinsic link between
the de Sitter radius and the Planck length via $\ell_{dS}^{2} / \ell_{pl}^{2} \sim N$.
In other words, the cosmological constant is always parametrically small in Planck units.
The finite number of degrees of freedom at finite $N$
is also suggestive of a holographic theory of de Sitter space \cite{Banks:2011av}.

\section*{Acknowledgements}

We thank N. Arkani-Hamed, N. Berkovits, S. Caron-Huot, T. Hartman, C. Hull, D. Karabali, M.
Kiermaier, J. Maldacena, V.P. Nair, D. Skinner, C. Vafa, E. Verlinde and E. Witten for
helpful discussions. The work of JJH is supported by NSF grant PHY-0969448 and
by the William Loughlin membership at the Institute for Advanced Study. The
work of HV is supported by NSF grant PHY-0756966


\bibliographystyle{titleutphys}
\bibliography{GMM}

\end{document}